\documentclass[fleqn,usenatbib]{mnras}

\usepackage{newtxtext,newtxmath}

\usepackage[T1]{fontenc}
\DeclareUnicodeCharacter{0327}{\c{}}
\DeclareRobustCommand{\VAN}[3]{#2}
\let\VANthebibliography\thebibliography
\def\thebibliography{\DeclareRobustCommand{\VAN}[3]{##3}\VANthebibliography}

\usepackage{graphicx}	
\usepackage{amsmath}	

\usepackage{multirow}
\usepackage{bm}
\newcommand{\tbabs}{\texttt{tbabs}}
\newcommand{\ebv}{$E(B-V)$}
\newcommand{\met}{$12 + \log$(O/H)}
\newcommand{\nism}{$n_\text{ISM}$}

\newcommand{\mout}{$\dot{m}_\text{ejec}$}

\newcommand{\rsph}{R$_{\text{sph}}$}

\newcommand{\rin}{R$_\text{in}$}
\newcommand{\rout}{$R_{\text{out}}$}
\newcommand{\vs}{$v_{\text{shock}}$}

\newcommand{\ecutoff}{$E_{\text{cutoff}}$}

\newcommand{\hst}{\textit{HST}}
\newcommand{\swift}{\textit{Swift}}
\newcommand{\xmm}{\textit{XMM-Newton}}
\newcommand{\nustar}{\textit{NuSTAR}}
\newcommand{\nh}{$N_\mathrm{H}$}

\newcommand{\theulx}{NGC~1313~X--2}
\newcommand{\paper}{\citet{gurpide_quasi_2024}}
\newcommand{\xspec}{\textsc{XSPEC}}
\newcommand{\cloudy}{\texttt{Cloudy}}

\newcommand{\ha}{H$\alpha$}
\newcommand{\hb}{H$\beta$}
\newcommand{\oiii}{[O~{\sc iii}]$\lambda$5007}
\newcommand{\heii}{He{\sc ii} $\lambda$4686}

\newcommand{\fluxcgs}{ergs~s$^{-1}$~cm$^{-2}$}
\newcommand{\lumcgs}{ergs~s$^{-1}$}

\title[The UV emission of the PULX NGC~1313~X--2]{Absence of nebular \heii\ constrains the UV emission from the Ultraluminous X-ray pulsar NGC~1313~X--2}

\author[A. G\'urpide et al.]{
A.~G\'urpide$^1$\thanks{E-mail: a.gurpide-lasheras@soton.ac.uk}, N.~Castro~Segura$^{2,1}$, R.~Soria$^{3,4,5}$ \& M.~Middleton$^1$
\\
$^{1}$School of Physics \& Astronomy, University of Southampton, Southampton, Southampton SO17 1BJ, UK \\
$^{2}$Department of Physics, University of Warwick, Gibbet Hill Road, Coventry CV4 7AL, UK\\
$^{3}$INAF-Osservatorio Astrofisico di Torino, Strada Osservatorio 20, I-10025 Pino Torinese, Italy  \\
$^{4}$Sydney Institute for Astronomy, School of Physics A28, The University of Sydney, Sydney, NSW 2006, Australia \\
$^{5}$College of Astronomy and Space Sciences, University of the Chinese Academy of Sciences, Beijing 100049, China
}

\date{Accepted XXX. Received YYY; in original form ZZZ}

\pubyear{2024}

\begin{document}
\label{firstpage}
\pagerange{\pageref{firstpage}--\pageref{lastpage}}
\maketitle

\begin{abstract} 
While much has been learned in recent decades about the X-ray emission of the extragalactic Ultraluminous X-ray sources (ULXs), their radiative output in the UV band remains poorly constrained. Understanding of the full ULX spectral energy distribution (SED) is imperative to constrain the accretion flow geometry powering them, as well as their radiative power. Here we present constraints on the UV emission of the pulsating ULX (PULX) NGC~1313~X--2 based on the absence of nebular {He{\sc ii} $\lambda$4686} emission in its immediate environment. To this end, we first perform multi-band spectroscopy of the ULX to derive three realistic extrapolations of the SED into the unaccessible UV, each predicting varying levels of UV luminosity. We then perform photo-ionization modelling of the bubble nebula and predict the {He{\sc ii} $\lambda$4686} fluxes that should have been observed based on each of the derived SEDs. We then compare these predictions with the derived upper limit on {\heii} from MUSE data, which allows us to infer a UV luminosity $L_\mathrm{UV} \lesssim 1 \times 10^{39}$ erg/s in the PULX NGC~1313~X--2. Comparing the UV luminosity inferred with other ULXs, our work suggests there may be an intrinsic difference between hard and soft ULXs, either related to different mass-transfer rates and/or the nature of the accretor. However, a statistical sample of ULXs with inferred UV luminosities is needed to fully determine the distinguishing features between hard and soft ULXs. Finally, we discuss ULXs ionising role in the context of the nebular {He{\sc ii} $\lambda$4686} line observed in star-forming, metal-poor galaxies.
\end{abstract}

   \maketitle
%
\section{Introduction} \label{sec:bubble_intro}

It is now established that super-Eddington accretion powers most of the extragalactic Ultraluminous X-ray sources (ULXs) discovered more than 30 years ago \citep{fabbiano_x_1989}. These are still empirically defined as point-like sources exceeding the classical Eddington limit for a 10 $M_{\sun}$ black hole (BH) ($L_\mathrm{X}\approx1\times$10$^{39}$ erg/s) \citep[although the exact luminosity threshold is loosely defined;][]{kaaret_ultraluminous_2017}, but most of them are now recognised to be an extension of X-ray binaries into the high-mass transfer regime \citep{fabrika_ultraluminous_2021}. Their extreme X-ray luminosities \citep[e.g.][]{israel_accreting_2017} together with the discovery of neutron star (NS) ULXs \citep[e.g.][]{bachetti_ultraluminous_2014, furst_discovery_2016, israel_discovery_2017}, blueshifted absorption lines \citep{pinto_resolved_2016, pinto_ultraluminous_2017} and hundred-parsec blown optical \citep{pakull_optical_2002, soria_ultraluminous_2021, gurpide_muse_2022, zhou_very_2022} and radio bubbles \citep{lang_radio_2007, berghea_detection_2020,soria_ultraluminous_2021} are all supporting of a picture where the strong radiation pressure due to a high mass-transfer rate drives powerful ($\sim$10$^{39-40}$ erg/s) radiatively-driven outflows from the accretion disk \citep{pinto_resolved_2016}. Thus, in a nutshell, ULXs act as laboratories for extreme accretion and feedback onto the environment \citep{pinto_ultra-luminous_2023}.

Despite such coherent picture establishing the majority of ULXs as a physical class, the exact accretion flow geometry powering ULXs remains debated. On the one hand it is known that strong magnetic fields ($>$10$^{12}$ G) reduce the cross-section for electron scattering \citep{basko_limiting_1976, mushtukov_maximum_2015}, thus increasing intrinsic Eddington luminosities. This makes invoking high-magnetic fields to explain ULXs extreme luminosities appealing. Moreover, the Eddington limit for an accretion column geometry may substantially differ from the spherical symmetry assumed in the classical Eddington limit \citep{basko_limiting_1976, kawashima_comptonized_2012}. But this picture presents some problems. Estimated accretion timescales in ULXs \citep[$\approx$10$^{5-6}$ yr;][]{pakull_300-parsec-long_2010, soria_ultraluminous_2021, gurpide_muse_2022, zhou_very_2022} suggest the strong mass-transfer rate should have buried the natal magnetic field of the NS \citep{middleton_neutron_2024}, while magnetars are generally found in isolate systems \citep{king_no_2019}. Additionally, it is unclear whether mass-loss in the disk would be quenched for strong magnetic fields, due to the disk being truncated before reaching Eddington luminosities \citep{chashkina_super-eddington_2017}, rendering the aforementioned bubbles difficult to explain. On the other hand, modest or relatively weak magnetic fields ($<$10$^{12}$ G) should lead to an accretion flow geometry similar to that envisioned by \citet{shakura_black_1973}, where an outflowing wind-cone develops in the disk. Scattering in the winds could explain the transient pulsations in NS-ULXs \citep{sathyaprakash_discovery_2019} or why many remain potentially unseen \citep{castillo_discovery_2020, gurpide_long-term_2021, mushtukov_pulsating_2021}.

Observationally, the X-ray spectra in ULXs is known to differ from those observed in Galactic BH X-ray binaries \citep{gladstone_ultraluminous_2009, middleton_challenging_2011}. Based on X-ray spectral hardness, ULXs are classified as hard or soft \citep{sutton_ultraluminous_2013} depending on the relative contribution of super-Eddington winds onto the line of sight \citep{middleton_spectral-timing_2015}. Supersoft ULXs are therefore an extension of these two categories which occur presumably at higher inclination and/or higher-mass transfer rates \citep{urquhart_optically_2016}. Because mass-transfer rate is tied to mass-outflow rate and hence optical thickness of the wind, increases in mass-transfer rates reduce the opening angle of the wind cone funnel \citep{king_masses_2009, kawashima_comptonized_2012} and increase the outflow photosphere \citep[e.g.][]{poutanen_supercritically_2007, abolmasov_optically_2009}. Therefore ULXs observed to transit through these spectral states are understood as changes in the opening angle of the funnel or the outflow photosphere responding to mass-transfer rate changes \citep{urquhart_optically_2016, feng_nature_2016, pinto_ultraluminous_2017, gurpide_discovery_2021}. But because the ultimate appearance of a ULX in X-rays depends on the relative position of the line of sight and the opening angle of the funnel, there is always an inherent degeneracy between mass-transfer rate and inclination in this classification. In other words, asserting whether two ULXs differ on their inclination or mass-transfer rate is hard to determine from X-ray spectroscopy. Moreover, differences between soft and hard ULXs may also be attributed to the nature of the accretor \citep{pintore_pulsator-like_2017, jithesh_broadband_2020, amato_ultraluminous_2023} or even to the strenght of the NS magnetic field \citep{gurpide_long-term_2021}.

A way forward is to study the luminosity of the accretion flow at other wavelengths. The UV band is presumably dominated by the photosphere of the optically thick wind \citep{poutanen_supercritically_2007, abolmasov_optically_2009}. This is observed, for instance, in the putative edge-on ULX SS 433 \citep{dolan_ss_1997}. Because the wind should emit more isotropically and it is also tied to the mass-transfer rate, measuring the luminosity of ULXs in the UV/EUV bands may offer means to break the aforementioned degeneracy between inclination and mass-transfer rate. While direct measurements are hindered by the absorption of UV photons beyond the Lyman alpha break, this can be accomplished by measuring the nebular EUV-sensitive \heii\ line, a tracer of high-energy photons as first recognised around the luminous BH LMC X--1 by \cite{pakull_detection_1986}. This line acts as a photon-counter and its brightness can be used to infer the number of photons emitted by the high-energy source in the $\approx$54 eV--0.1 keV band. At higher energies, He+ becomes optically thin because the absorption cross-section falls off as $\nu^{-3}$, where $\nu$ is the photon frequency. Observations of this line around ULXs have been used to infer that ULXs are also Ultraluminous in the UV band \citep{abolmasov_kinematics_2008, abolmasov_optically_2009, kaaret_direct_2010}. But the number of ULXs with measured UV luminosities at present remains small, and a clear picture cannot yet emerge.

While the UV brightness of ULXs can have strong implications on the nature of the compact object and the accretion flow geometry powering them, it is also of importance to understand their capacity to ionize their environment. In particular, ULXs have recently received attention as a potential energy source to explain the so-called `nebular \heii\ problem'. This line is observed in the integrated spectra of some star-forming galaxies (i.e. AGN-less galaxies), where it is known `regular' stars do not produce enough UV photons to account for it \citep[see][and references therein]{shirazi_strongly_2012}. Wolf-Rayet (WR) stars can account for its presence in some galaxies, but the \heii\ line is often observed in the absence of WR signatures. Moreover, the strength of the \heii\ line is observed to increase in low-metallicity galaxies, where WR stars are expected to be less abundant \citep{shirazi_strongly_2012}. Given that \textit{some} ULXs are known to produce strong He{\sc iii} regions \citep{pakull_optical_2002, kaaret_photoionized_2009} and that they are more prevalent in low-metallicity galaxies \citep[e.g.][]{kovlakas_census_2020}, numerous studies have considered whether ULXs could be the missing source of high-energy photons in metal-poor star-forming galaxies \citep{simmonds_can_2021, kovlakas_ionizing_2022, garofali_modeling_2023}. The lack of knowledge of the ULX spectral energy distribution (SED) in the UV makes it difficult to ascertain whether that could be the case. 

In this paper, we turn the problem around and use the \textit{absence} of the EUV-sensitive \heii\ line around the ULX \theulx\ to put an upper limit on its EUV luminosity. This ULX is one of the few confirmed pulsating ULXs \citep[PULXs;][]{sathyaprakash_discovery_2019} and its X-ray spectrum is among the hardest in the ULX population \citep{sutton_ultraluminous_2013, pintore_pulsator-like_2017, gurpide_long-term_2021}, which could be interpreted as the source being viewed close to face-on \citep{sathyaprakash_discovery_2019}, or as the signature of a NS accretor \citep{gurpide_long-term_2021, amato_ultraluminous_2023}, or both. Most remarkably, the source is embedded in a $\approx$ 450 pc shock-ionized optical bubble \citep{pakull_optical_2002, ramsey_optical_2006, zhou_very_2022}, which is too large compared to supernova remnants \citep[e.g.][]{pakull_optical_2002}. Instead, its size and expansion velocity suggest a continuous outflow with mechanical power $P_\mathrm{w}$ $\approx$ 10$^{39}$ erg/s over $t\approx 10^6$ yr is responsible for it \citep{zhou_very_2022}. 

Here we will show that the inferred \heii\ upper limit suggest the UV luminosity in \theulx\ is likely below $\sim$1$\times$10$^{39}$ erg/s. We compare this measurement with the luminous ($L_\mathrm{UV} \sim$10$^{40}$ erg/s) soft ULX NGC~6946~X--1 \citep{abolmasov_optically_2009} and suggest only soft ULXs may produce enough He+ photons to explain the \heii\ line in star-forming, metal-poor galaxies. Hard ULXs seem instead, from the limited sample, to posses dimmer UV luminosities incapable of accounting for the \heii\ problem.

This paper is organised as follows: in Section~\ref{sec:bubble_data_reduction} we present the Multi-Unit Spectroscopic Explorer \citep[MUSE;][]{bacon_muse_2010} observations probing the nebular emission and the multi-band spectroscopic data we use to characterise the source SED.  Section~\ref{sec:multiband} presents the spectral analysis of the data, where we derive three possible extrapolations to the UV compatible with the optical and X-ray spectroscopic data. In Section~\ref{sec:cloudy} we present photo-ionization modelling of the nebula based on these three SEDs and rule out those producing UV luminosities incompatible with our upper limits on the \heii\ line derived from the MUSE observations. Finally, in Section~\ref{sec:discussion} we discuss our results and in Section~\ref{sec:conclusion} we present our conclusions.

\section{Data reduction}\label{sec:bubble_data_reduction}

\subsection{Multi-Unit Spectroscopic Explorer}\label{sub:muse_data}
\theulx\ was observed in four instances by MUSE (PI F.P.A. Vogt), all of them in seeing-limited Wide Field Mode (WFM), with the extended option that provides coverage down to $\lambda \sim$ 465\,nm. The details of these observations are given in Table~\ref{tab:data} and were recently analysed in \citet{zhou_very_2022}. The data was obtained from the ESO Science Portal\footnote{\url{https://archive.eso.org/scienceportal/home}}.
\begin{table*}
    \centering
    \caption{Log of the observations used in this work.}
    \label{tab:data}
    \begin{tabular}{ccccccc}
    \hline
     \hline
     \noalign{\smallskip}
        Telescope & Detector & ObsID & Date & Band & Exposure & Seeing$^a$ \\ 
         &  &  & yy-mm-dd & \AA & ks & \arcsec \\
         \hline
         \noalign{\smallskip}
\hst\ & WFCP2/PC/F555W & hst\_11227\_01\_wfpc2\_f555w$^b$ & 2009-11-15 & 4661--6224 & 1.1 & -- \\
\noalign{\smallskip}
\hst\ & ACS/HRC/F330W & hst\_9796\_a2\_acs\_hrc\_f330w & \multirow{4}{*}{2003-11-22} &  3158--3567 & 2.76 &  -- \\ 
\hst\ & ACS/WFC/F435W & hst\_9796\_02\_acs\_wfc\_f435w &  &  3976--4680 & 2.52 &  -- \\ 
\hst\ & ACS/WFC/F555W & hst\_9796\_02\_acs\_wfc\_f555w &  &  4936--5784 & 1.16 &  -- \\ 
\hst\ & ACS/WFC/F814W & hst\_9796\_02\_acs\_wfc\_f814w &  &  7277--8817 & 1.16 &  -- \\ 
\noalign{\smallskip}
\hst\ & ACS/SBC/F140LP & hst\_14057\_01\_acs\_sbc\_f140lp &  \multirow{7}{*}{2015-12-05} &  1384--1671 & 1.20 &  -- \\ 
\hst\ & WFC3/UVIS/F225W & hst\_14057\_a1\_wfc3\_uvis\_f225w &  &  2162--2570 & 1.44 &  -- \\
\hst\ & WFC3/UVIS/F336W & hst\_14057\_a1\_wfc3\_uvis\_f336w &  &  3168--3541 & 1.02 &  -- \\
\hst\ & WFC3/UVIS/F438W & hst\_14057\_a1\_wfc3\_uvis\_f438w &  &  4697--5914 & 1.05 &  -- \\
\hst\ & WFC3/UVIS/F555W & hst\_14057\_a1\_wfc3\_uvis\_f555w &  &  5047--5564 & 1.17 &  -- \\ 
\hst\ & WFC3/UVIS/F814W & hst\_14057\_a1\_wfc3\_uvis\_f814w &  &  7266--8829 & 1.74 &  -- \\
\hst\ & WFC3/IR/F125W & hst\_14057\_a1\_wfc3\_ir\_f125w &  &  11466--13506 & 1.80 &  -- \\ 
\noalign{\smallskip}
\hst\ & ACS/SBC/F140LP & hst\_14057\_02\_acs\_sbc\_f140lp & \multirow{7}{*}{2016-03-23} &  1384--1671 & 1.20 &  -- \\ 
\hst\ & WFC3/UVIS/F225W & hst\_14057\_a2\_wfc3\_uvis\_f225w &  &  2162--2570 & 1.44 &  -- \\ 
\hst\ & WFC3/UVIS/F336W & hst\_14057\_a2\_wfc3\_uvis\_f336w &  &  3168--3541  & 1.02 &  -- \\ 
\hst\ & WFC3/UVIS/F438W & hst\_14057\_a2\_wfc3\_uvis\_f438w &  &  4093--4557 & 1.05 &  -- \\ 
\hst\ & WFC3/UVIS/F555W & hst\_14057\_a2\_wfc3\_uvis\_f555w &  &  4697--5914 & 1.17 &  -- \\ 
\hst\ & WFC3/UVIS/F814W & hst\_14057\_a2\_wfc3\_uvis\_f814w &  &  7266--8829 & 1.74 &  -- \\
\hst\ & WFC3/IR/F125W & hst\_14057\_a2\_wfc3\_ir\_f125w &  &  11466--13506 & 1.80 &  -- \\ 
\hline
\noalign{\smallskip}
 VLT & MUSE& ADP.2019-11-29T11:13:19.082  & 2019-10-16  &  4650-9350  & 1.4   & 1.37--1.09\\
 VLT & MUSE & ADP.2019-12-20T09:12:08.137 & 2019-11-12  &  4650-9350 & 1.4   & 0.51--0.51\\
 VLT & MUSE & ADP.2019-12-20T09:12:08.101 & 2019-11-12  &  4650-9350  & 1.4   & 0.59--0.53\\
 VLT & MUSE & ADP.2019-12-20T09:12:08.153 & 2019-11-12 &  4650--9350  & 1.4   & 0.55--0.54\\
          \noalign{\smallskip}
         \hline
         \hline
    \end{tabular}
    \begin{minipage}{0.99\linewidth}
 \textbf{Notes}.$^a$Seeing at the start and end of the night.\\
  $^b$Observation only used for the astrometric correction of the MUSE datacube (see text for details).\\
\end{minipage}
\end{table*}

We inspected each of the cubes individually using \texttt{ds9} \citep{joye_new_2003} as well running the analysis we describe below in each cube. The sky during the three November observations (all taken during the same night) was partially cloudy. We found these datacubes to have very poor signal-to-noise ratio (S/N), with most of the nebular emission lines detected well below a S/N of 5 throughout the entire cube. Despite the better seeing (Table~\ref{tab:data}), we suspect the poor weather conditions may have compromised the data quality during these observations. In contrast, the sky was clear during the single October observation. We found this datacube to have a much higher S/N ratio, allowing us to detect fainter lines such as \oiii\ even in the fainter eastern part of the nebula. We therefore focused on this cube for our analysis and refer to it as cube 1 hereafter. 

\begin{figure}
    \centering
    \includegraphics[width=0.49\textwidth]{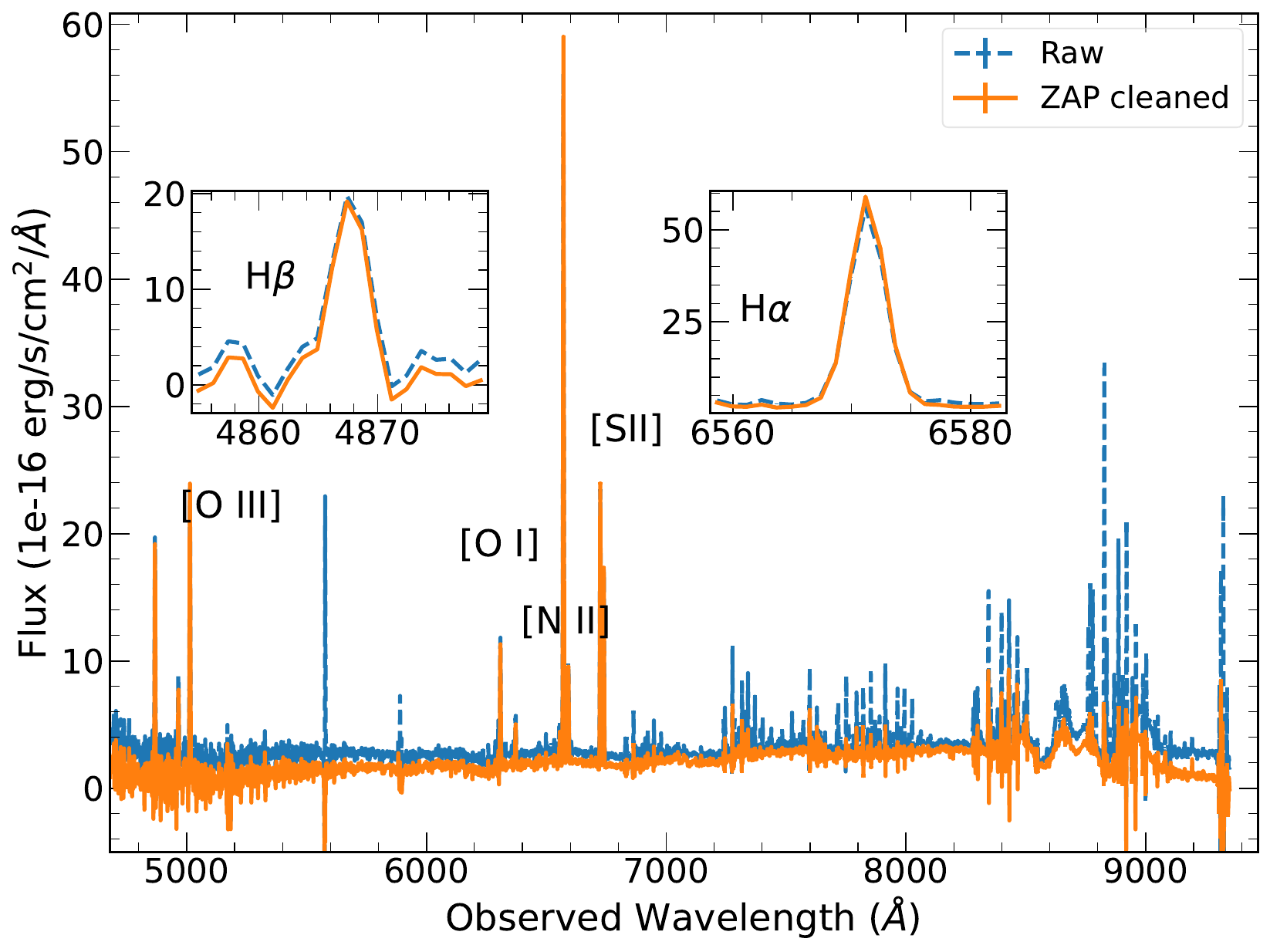}
    \caption{Integrated spectrum of the bubble before (blue dashed line) and after (orange solid line) applying the sky correction using \texttt{ZAP} \citep{soto_zap_2016}. Important nebular lines are labelled. The inset panels show a close-up look at the H$\beta$ (left) and H$\alpha$ lines. The correction is minor around these lines, indicating the source signal has not been affected by the correction (see text for details).}
    \label{fig:zap_bubble}
\end{figure}

Given the sparse field around the target, we opted to clean further the data cubes from telluric features using the Zurich Atmosphere Purge \citep[ZAP;][]{soto_zap_2016}. We masked the bubble and the brightest sources in the field to obtain a clean sky solution. We set the window used to remove the continuum features to 30 pixels to trace more closely the source signal \citep{soto_zap_2016}. One of the main parameters affecting the subtraction of the sky residuals is the number of eigenvectors used in the principal component analysis used to reconstruct the residual sky spectrum. Higher values tend to result in a more aggressive correction, potentially subtracting some of the astronomical signal. By default the task attempts to select an optimal number of eigenvalues by looking at the reduction in variance in the datacube after the residual sky spectra is subtracted as more eigenvectors are added. The eigenvalue at which the reduction in variance starts to plateau or enters the linear regime (at which point the source signal might start to be removed) is considered to be the optimal value. We tested different eigenvalues and for each eigenvalue-corrected cube we extracted an integrated spectrum of the bubble (Figure~\ref{fig:zap_bubble}) and judged the quality of the correction visually from this spectrum. We found that for the optimally-calculated value of 52, the task removed some of the source signal, particularly visible in a reduction of in flux of some of the most prominent nebular lines. We visually inspected the decrease in variance as a function of the number of eigenvector as suggested by the authors and found that the values around 10--15 seemed more appropriate, where the loss in source signal was nearly absent. We finally settled for 10 eigenvectors. 

Figure~\ref{fig:zap_bubble} shows the integrated spectrum of the bubble surrounding \theulx\ (details on the spectrum extraction are provided below) before and after the sky correction. As can be seen, the correction was more pronounced upwards of 7000 \AA, where telluric absorption is more important. For the strong nebular emission lines in which we are interested here, the correction had mostly the effect of slightly improving their S/N by affecting mostly the continuum.

The cubes were then aligned to an \hst\ reference image. Following \citet{gurpide_muse_2022}, we chose an image in the F555W filter owing to the substantial overlap with the MUSE wavelength range. We retrieved a calibrated Wide Field Planetary Camera 2 image of the field around \theulx\ from the Hubble Legacy Archive \footnote{\url{https://hla.stsci.edu/}}. The details of this observation are also given in Table~\ref{tab:data}. To obtain the position of \theulx\ in the datacube, following \citet{gurpide_muse_2022} we first fitted a 2D circular Gaussian profile to the \hst\ counterpart. We obtained a FWHM 0.103$\pm$0.002\arcsec\ (1$\sigma$ uncertainty). Adding this value in quadrature to the MUSE pixel scale (0.2\arcsec), we obtained a 3$\sigma$ uncertainty on the position of \theulx\ of $\sim$0.67\arcsec.

\begin{figure}
    \centering
    \includegraphics[width=0.49\textwidth]{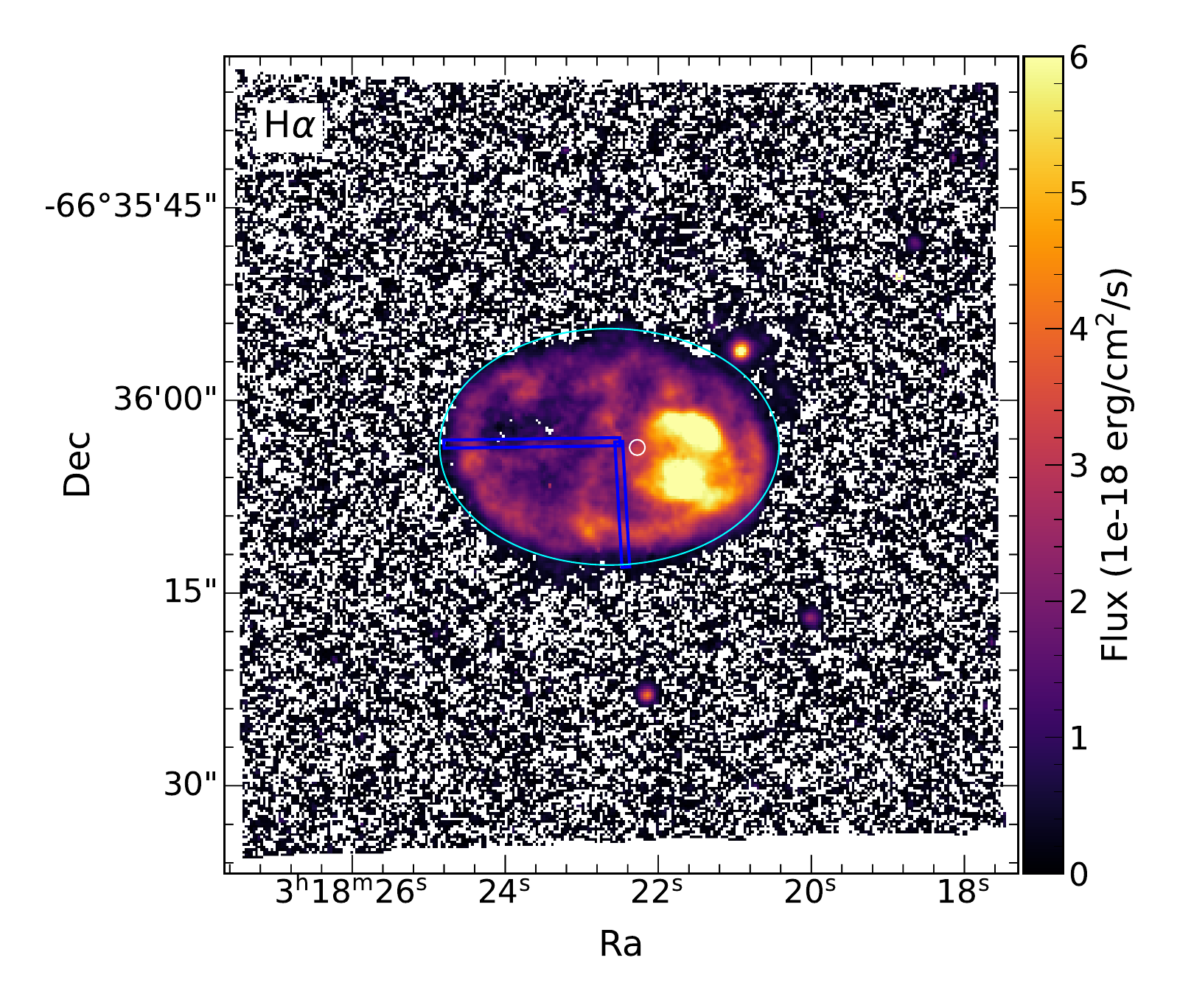}
    \caption{The bubble nebula around \theulx\ in \ha\ emission. Pixels below a S/N below 3 have been masked. The 3$\sigma$ uncertainty on the position of \theulx\ is shown as a white circle while the cyan ellipse shows the region used to extract the integrated spectrum shown in Figure~\ref{fig:zap_bubble}, from which we derive the upper limit on the \heii\ line in Section~\ref{sub:heii}. The rectangular blue boxes show the profiles extracted to determine the density profile of the bubble (Section~\ref{sec:cloudy}). North is up and East is left.}
    \label{fig:bubble_ha}
\end{figure}

Figure~\ref{fig:bubble_ha} shows an \ha\ image of the bubble surrounding \theulx. The image was extracted using pixel-by-pixel Gaussian line fitting with CAMEL \citep{epinat_massiv_2012, epinat_ionised_2018}. The position of \theulx\ is indicated by a white circle and the region used to extract the bubble spectrum in Figure~\ref{fig:zap_bubble} as a cyan ellipse. A full analysis of the bubble will be presented in a forthcoming publication. Here we focus on the constraints we can infer on the emission of \theulx\ from the non-detection of the \heii\ line in its vicinity.

\subsubsection{Upper limits on the \heii\ line} \label{sub:heii}
\begin{figure}
    \centering
    \includegraphics[width=0.49\textwidth]{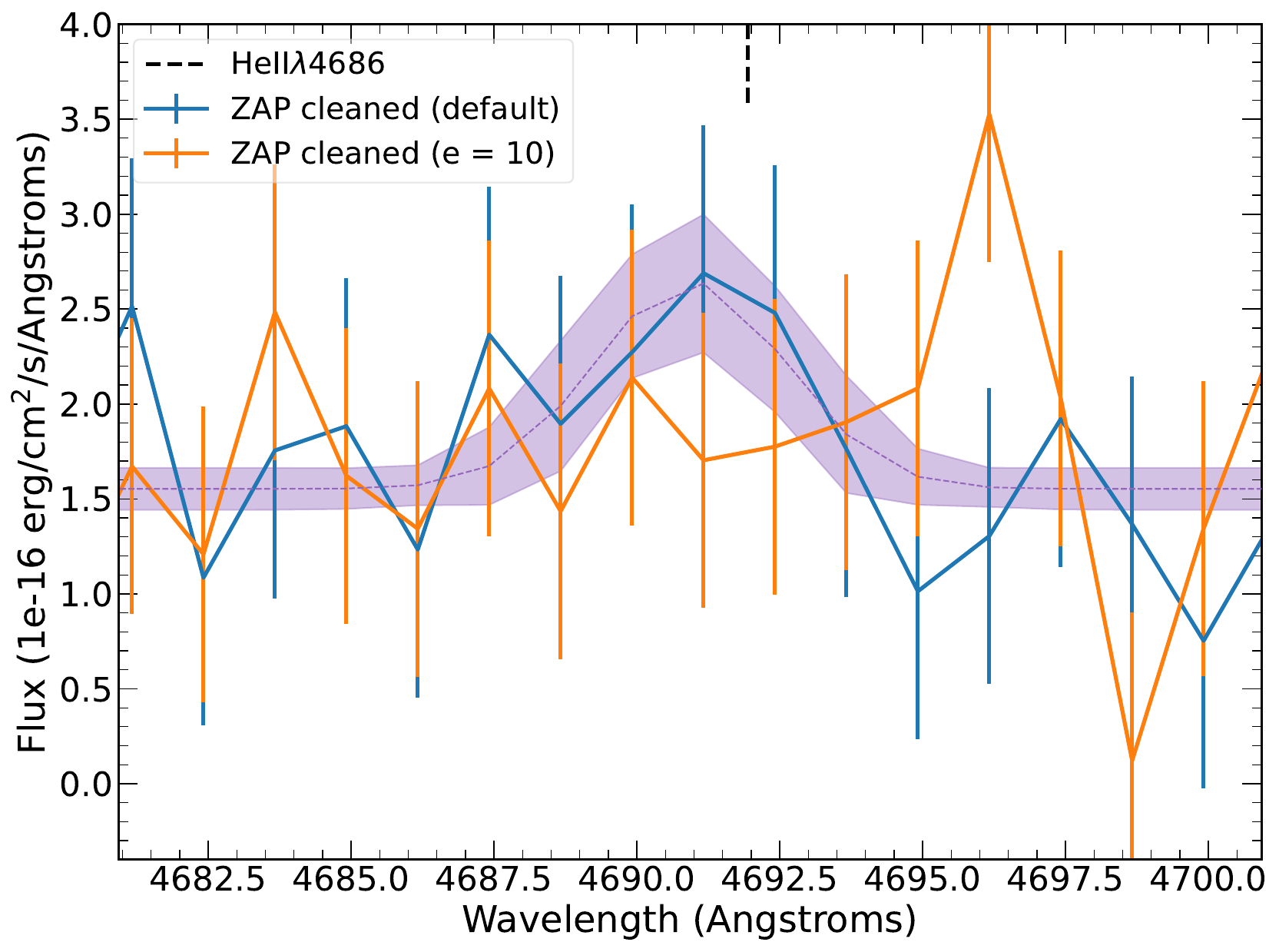}
    \caption{Best-fit to the putative \heii\ line of the integrated spectrum of the bubble. Two spectra are shown corrected for sky residuals using ZAP with two different settings (see text for details). The dashed black line shows the expected position of the line based on the heliocentric velocity measured by \citet{ramsey_optical_2006} (380~km/s). The best-fit model (Gaussian + Constant) and its 1$\sigma$ uncertainty to the putative line are shown as a magenta dashed line and corresponding shaded area. The line is consistent with being null at the 3$\sigma$ level and therefore it is considered an upper limit.}
    \label{fig:heII}
\end{figure}

We did not find the \heii\ line to be obviously present in any region around the ULX. We therefore inspected the integrated spectrum of the bubble (Figure~\ref{fig:zap_bubble}) in an attempt to detect the faint \heii\ line. Figure~\ref{fig:heII} shows a close-up of the bubble spectrum (Figure~\ref{fig:zap_bubble}) around the \heii\ line (orange solid line). There is no obvious feature at the expected position of the line considering the heliocentric velocity of 380~km/s derived by \citet{ramsey_optical_2006}. The non-detection is in agreement with the non-detection also reported by \citep[][see also \citet{zampieri_ultraluminous_2004}]{ramsey_optical_2006}. Forcing a Gaussian line at the expected line position onto the continuum -- modelled with a constant\footnote{Equivalent results were found using a linear function for the continuum.} -- we found the line to be detected below the 1$\sigma$ detection level. Correcting for foreground extinction (see Section~\ref{sub:extinction}), we put a 1$\sigma$ upper limit on its flux of $F($\heii) $< 4\times10^{-16}$ \fluxcgs\ or $L($\heii) $< 8\times10^{35}$ \lumcgs at a distance of 4.25~Mpc. 

We have also investigated whether using a more aggressive correction (i.e. using the default values from \texttt{ZAP} as outlined in Section~\ref{sub:muse_data}) would reveal any underlying feature covered by noise. This spectrum is also shown in Figure~\ref{fig:zap_bubble} (blue solid line). There is very marginal evidence for the line being present close to the expected position, but the 3 $\sigma$ uncertainty on the flux is still consistent with zero. Therefore owing to the uncertainties associated with the correction, position, and flux of the line we put a more conservative 3$\sigma$ upper limit on its presence of $F$(\heii) $< 6\times10^{-16}$ \fluxcgs\ or $L$(\heii) $< 1\times10^{36}$ \lumcgs. This upper limit is likely conservative as there is a non-negligible chance that we are just fitting noise, but nevertheless it is compatible with the estimate given above. 

For comparison, this upper limit is a an order of magnitude lower than the integrated luminosity of \heii\ in the NGC 6946 X-1 nebula \citep{abolmasov_optical_2008} and suggest the line luminosity could be comparable to that observed in NGC~1313~X--1 ($\sim$7.5--9$\times10^{35}$ \lumcgs;\paper). This already points to a situation in which the EUV emission in \theulx\ is not particularly bright. In Section~\ref{sec:cloudy} we provide a quantitative estimate using photo-ionization modelling.
\subsection{\swift-XRT}
\begin{figure*}
    \centering
    \includegraphics[width=0.98\textwidth]{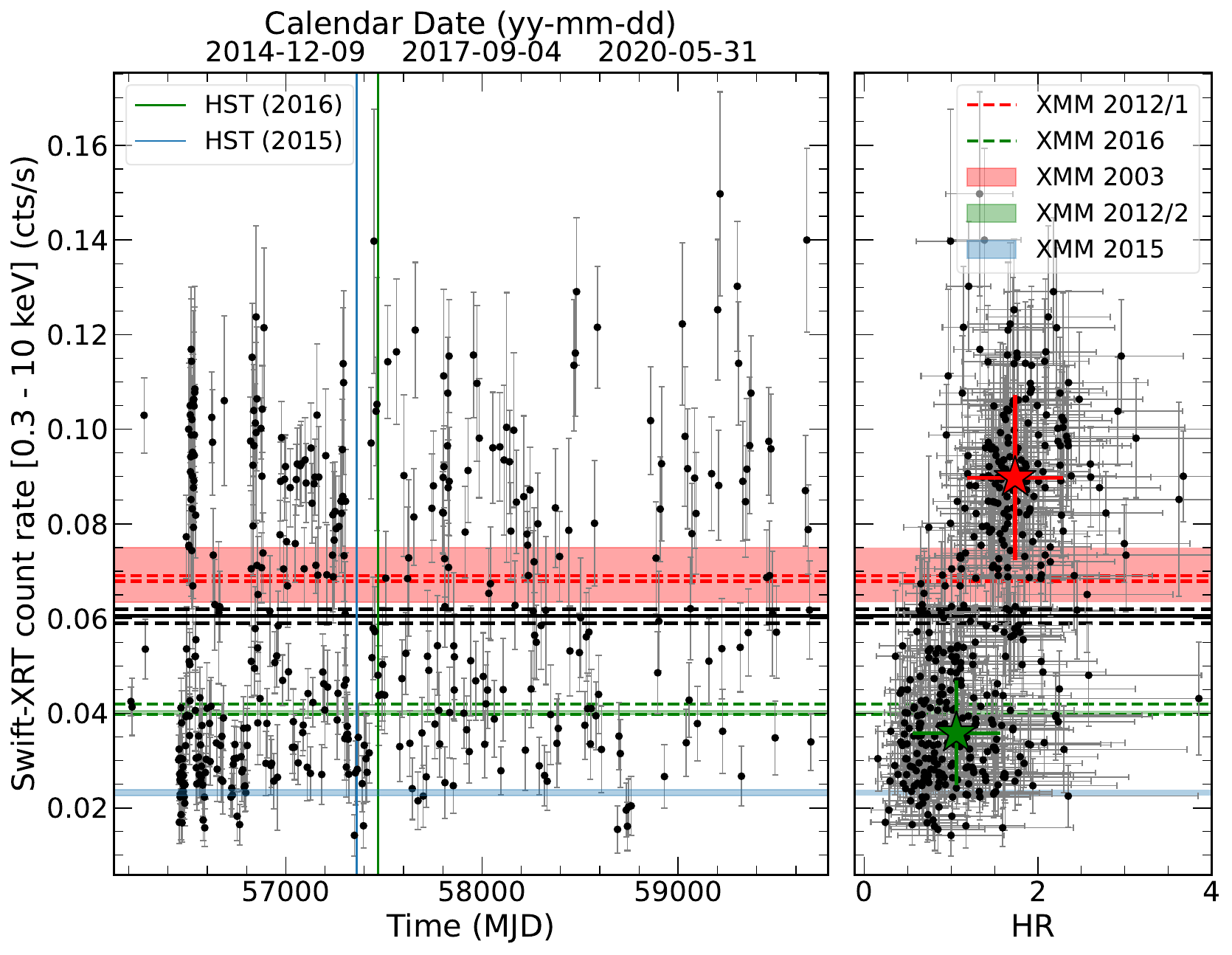}
    \caption{\swift-XRT lightcurve and corresponding hardness ratio showing the variability and spectral states of NGC~1313~X--2. (Left) \swift-XRT lightcurve binned by observations. The solid and dashed lines indicate the mean count rate and its standard error. The \hst\ observations from 2015 and 2016 are shown as vertical blue lines. The horizontal shaded areas (or dashed lines) indicate the different \xmm\ observations flux levels converted to \swift-XRT equivalent count rates. Both the simultaneous \xmm\ observation to the \hst\ observations are shown as well as the dataset used to characterise the spectral states (see text for details). (Right) Hardness-intensity ratio. The hardness is defined as count rate in the 1.5 --10 keV band over the count rate in the 0.3 -- 1.5 keV band. The green and red stars show the mean and and spread of the high and low states of NGC~1313~X--2, whose threshold has been set at 0.06ct/s. Low SNR datapoints with HR >4 have been omitted for clarity. }
    \label{fig:swift_hst}
\end{figure*}

As for \citet{gurpide_quasi_2024, sathyaprakash_multi-wavelength_2022}, we take advantage of the long baseline offered by the \swift-XRT \citep{burrows_swift_2005} to characterise the source spectral states. We reduced all available \swift-XRT observations using the standard online tools \citep{evans_online_2007, evans_methods_2009}, considering any observations with a detection significance above 2$\sigma$. Regular monitoring of the source started in $\sim$2012 with a median cadence of $\sim$5 days. We show this section of the lightcurve in Figure~\ref{fig:swift_hst} along with the hardness intensity diagram (defined as the count rates in the 1.5--10 keV and over the count rates in the 0.3 -- 1.5 keV). The well-documented bimodal behaviour of the source \citep{weng_evidence_2018, sathyaprakash_multi-wavelength_2022} is apparent in the Figure. The bright and dim states -- hereafter high and low states \citep[see also][]{sutton_irradiated_2014} -- are illustrated by a red and green stars respectively on the HR diagram. These have been calculated by taking the mean of the observations above and below 0.06 ct/s (the mean count rate), respectively.

\subsection{\xmm\ and \nustar}

Based on the epochs of the \hst\ observations (Section~\ref{sub:hst}), we selected (nearly) simultaneous \xmm\ observations to characterise the X-ray spectral state at the time of the \hst\ observations (Section~\ref{sub:hst}). The first set of nearly simultaneous \hst\ and \xmm\ multi-band analysis was presented in \citet{sutton_irradiated_2014}, corresponding to obsid 0150280101 taken in 2003-11-25, close to the 2003-11-22 \hst\ observations. {\theulx} can vary in timescales shorter than the 3-day time interval between the X-ray and optical observations \citep[][and Figure~\ref{fig:swift_hst}]{grise_ultraluminous_2008, gurpide_long-term_2021}, but variations in the X-ray flux are generally smaller than our defined high and low states \citep[see e.g.][]{robba_broadband_2021}, while \theulx\ often remains stable during these short periods. While we obviously cannot rule out that \theulx\ has varied in this time span, this \hst\ dataset is the closest to high-state observations in the archives. Therefore, owing to the lack of strictly simultaneously \hst\ observations probing the high state, we proceeded under the assumption that \theulx\ was at the same X-ray flux as observed by \xmm\ during the \hst\ observations (we discuss this assumption in more detail below). The other existing \hst\ and \xmm\ multi-band dataset was presented in \citet{sathyaprakash_multi-wavelength_2022}, for which we selected obsid 0764770101 taken in 2015-12-05 and obsid 0764770401 taken in 2016-03-23, both with dedicated coordinated \hst\ observations.

We next investigated the X-ray flux of the source at the time of the \hst\ observations to place the \hst\ observation into one of the the two spectral states defined above (Figure~\ref{fig:swift_hst}). \xmm\ spectral products were taken from \citet{gurpide_long-term_2021} and the three EPIC cameras were rebinned using the scheme proposed by \citet{kaastra_optimal_2016}. Spectral fitting was carried out using \xspec\ version 12.13.1 \citep{arnaud_xspec:_1996} over the band where the source dominated above the background (typically 0.3 to 8--9 keV). The spectra had enough counts to assume Gaussian statistics, so the $\chi^2$ was employed for minimisation.

We fitted all datasets with a simple absorbed cutoff powerlaw to determine the observed flux and equivalent \swift-XRT count rate, which we determined by convolving the best-fit models with the latest \swift-XRT response files. Absorption was taken into account using the \texttt{tbabs} model \citep{wilms_absorption_2000} with cross-sections of \citet{verner_atomic_1996}. Best-fit parameters, observed flux and \swift-XRT equivalent count rates are given in Table~\ref{tab:xmm_states}. This simple model provided a good description of all continua except for obsid 0764770101 ($\chi^2 = 364$/270), which caught the source at its lowest flux and instead showed a soft excess on top of the powerlaw. For this observation we therefore employed a \texttt{bbody} and a \texttt{powerlaw}.
\begin{table*}
    \centering
    \caption{Log of \xmm\ observations used to determine the spectral states of \theulx\ during the \hst\ observations.}
    \label{tab:xmm_states}
     \resizebox{\textwidth}{!}{\begin{tabular}{ccccccccc} 
    \hline
    \hline
    \noalign{\smallskip}
 Obsid    & Date       & \nh\                & $\Gamma$ & $E_\mathrm{cutoff}$/$kT^a$ &  $\chi^2$/dof & Flux         &  \swift-XRT count rate & State  \\
          &     yyyy-mm-dd     & 10$^{20}$ cm$^{-2}$ &           & keV                &               &  10$^{-12}$ \fluxcgs\ & ct/s &  \\
        \hline
 \noalign{\smallskip}
0764770101  & 2015-12-15 & 20$\pm2$ & 1.75$\pm0.07$ & 0.22$\pm0.02$ &  230/269 & 0.83$\pm0.02$ & 0.0232$\pm0.0006$ & Very Low \\
        \noalign{\smallskip}
0693851201 & 2012-12-22 & $25\pm2$ &1.3$\pm0.1$ & 2.3$\pm0.2$  & 305/263 & 1.13$\pm0.01$ & 0.0401$\pm0.0004$ & \multirow{2}{*}{Low} \\
0764770401 &  2016-03-23  & 25$\pm4$ &  1.3$\pm0.2$ &  2.2$_{-0.4}^{+0.5}$ & 237/217 & 1.16$\pm0.03$&  0.0409$\pm0.001$ &  \\
  \noalign{\smallskip} 
0150280101 & 2003-11-25 & 21$_{-8}^{+9}$ &  0.9$\pm0.6$ & 3$_{-1}^{+3}$    & 161/166 & 2.4$\pm0.2$ & 0.0688$\pm0.006$ &   \multirow{2}{*}{High}\\
0693850501 & 2012-12-16 & 22$\pm1$ & 0.88$\pm0.07$ & 2.5$_{-0.1}^{+0.2}$   & 323/308 & 2.37$\pm0.02$ &  0.0685$\pm0.0006$ & \\
  \noalign{\smallskip}
  \hline
  \hline
    \end{tabular}}
        \begin{minipage}{\linewidth}
     $\textbf{Notes.}$  Uncertainties at the 90\% confidence level. \\
     $^a$We employed a cutoff powerlaw for all observations except for obsid 0764770101, for which we employed a simple powerlaw and a blackbody.
\end{minipage}
\end{table*}

Based on the flux levels and \swift-XRT equivalent count rates we can see that observation 0150280101 (from 2003) probes the high state, while obsids 0764770101/401 (from 2015 and 2016 respectively) probe instead epochs of lower fluxes. Since our aim is to determine a spectrum close to the temporal weighted average (showed as a black solid line), we considered the observations from the high state (2003) as well as the observations from the low state from 2016 (where the source is in the low state, but brighter than in 2015). 

Based on these observations and the analysis presented in \citet{gurpide_long-term_2021}, we then retrieved longer, higher-quality observations on the same flux/hardness levels as the 2003 and 2016 epochs to characterise the broadband SED (Section~\ref{sec:multiband}).We thus retrieved obsids 0693851201 and 0693850501 for the low and high states respectively. Their fit parameters are also reported in Table~\ref{tab:xmm_states} and are fully consistent with the obsids from the corresponding states. The equivalent \swift-XRT count rates of all these observations are shown in Figure~\ref{fig:swift_hst} to illustrate their agreement and the source flux level they probe in comparison to the source overall variability.

Finally, we looked for simultaneous \nustar\ observations to extend the coverage to higher energies. We retrieved the coordinated \nustar\ obsids 30002035002 and 30002035004 for the high and low states respectively. These were reprocessed using the \texttt{nustardas} software version 1.9.7 with \textsc{CALDB} version 20230223 and spectral products were extracted as detailed in \citet{gurpide_long-term_2021}. These were rebinned as for the EPIC data and analysed over the band were the source dominated over the background (typically the 3--14 keV band).

\subsection{\textit{Hubble Space Telescope}} \label{sub:hst}

As stated above, the multi-band \hst\ and X-ray analysis of \theulx\ have been presented in \citet{sutton_irradiated_2014} and \citet{sathyaprakash_multi-wavelength_2022}. The observations presented in \citet{sutton_irradiated_2014} were taken on the 22nd of November of 2003 and were nearly simultaneous with \xmm\ obsid 0150280101 (Table~\ref{tab:xmm_states}). As we have shown above, these observations probe a state close to the high state of \theulx\ (Figure~\ref{fig:swift_hst}). The set of observations presented in \citet{sathyaprakash_multi-wavelength_2022} probed instead two states of lower flux, of which we only consider the brighter one (the 2016 observations; Figure~\ref{fig:swift_hst}). 

Our data reduction mirrors that followed in \citet{gurpide_quasi_2024}. We downloaded all \hst\ archival data available from the MAST portal \footnote{\url{https://mast.stsci.edu/portal/Mashup/Clients/Mast/ Portal.html}}, which provides calibrated, geometrically-corrected drizzled images. We performed aperture photometry of \theulx\ using a circular region of 0.2\arcsec\ in radius for the ACS/WFC, WFC3/UVIS and ACS/HRC detectors (corresponding to 4, 5 and 8 pixels, respectively) and 0.4\arcsec\ for the ACS/SBC and WFC3/IR (corresponding to $\approx$13 pixels and 3 pixels, respectively) centroided on the source position using 2D Gaussian fitting. In order to correct the counts for the finite aperture, we again followed a two step process: first, we looked for isolated, bright, unsaturated stars from the same chip as \theulx, to estimate the correction to a 10 pixel aperture. We typically used 1 to 5 stars per filter. Due to the smaller field of view and lack of bright UV stars, this was not possible for the ACS/SBC for which we simply relied on the tabulated values \citep[see also][]{sathyaprakash_multi-wavelength_2022}. We then corrected the 10-pixel-radius count rates derived for all images to an infinite aperture using the tabulated values. 

The background level was estimated as per \citet{gurpide_quasi_2024}, selecting a source-free region and using the 3$\sigma$-clipped median count rate and its standard deviation as the background estimate. The uncertainties on the final background-subtracted count rates were derived assuming Poisson statistics for the source and background regions and accounting for the uncertainty on the aperture correction.

\begin{table*}
    \centering
    \caption{Extinction-corrected fluxes for the counterpart of \theulx. An absorbed powerlaw was employed in all instances for the conversion between count rates and (intrinsic) fluxes.}
    \label{tab:hst_fluxes}
    \resizebox{\textwidth}{!}{\begin{tabular}{cccccccc}
\hline \hline 
 \noalign{\smallskip} 
Date       & Filter & Pivot & State & $\alpha$ & \ebv$^a$ & $\chi^2$/dof &  Flux \\ 

yyyy-mm-dd &         &  \AA &       &          &  mag  &              & 10$^{-18}$ erg/s/cm$^2$/\AA \\ 
\noalign{\smallskip}
\hline
\noalign{\smallskip}
\multirow{4}{*}{2003-11-22} & ACS/HRC/F330W & 3362.8 & \multirow{4}{*}{High} &  \multirow{4}{*}{3.4$\pm0.1$}  &  \multirow{4}{*}{0.057$\pm$0.03}  &  \multirow{4}{*}{0.95 / 1}  &  10.04$\pm$1.3  \\ 
& ACS/WFC/F435W & 4329.1 &  &    &   &   &  4.3$\pm$0.4         \\ 
& ACS/WFC/F555W & 5360.2 &  &    &    &    &  2.09$\pm$0.18     \\ 
& ACS/WFC/F814W & 8047.3 &  &   &   &    &  0.53$\pm$0.03       \\ 
\noalign{\smallskip}
\hline
\noalign{\smallskip}
\multirow{6}{*}{2016-03-24} &  WFC3/UVIS/F336W & 3354.7 & \multirow{6}{*}{Low} &  \multirow{4}{*}{3.24$\pm0.08$}  &  \multirow{5}{*}{0.05$\pm0.03$}  &  \multirow{4}{*}{4.05 / 2}  &  10.2$\pm$1.2  \\
& WFC3/UVIS/F438W & 4325.1 &  &    &  &  &  4.5$\pm$0.4 \\
& WFC3/UVIS/F555W & 5307.9 &  &    &  &  &  2.3$\pm$0.2  \\ 
& WFC3/UVIS/F814W & 8029.3 &  &    &   &    &  0.60$\pm$0.03  \\ 
& WFC3/IR/F125W & 12486.1 &  &    &    &    &  0.140$\pm$0.001  \\ 
\noalign{\smallskip}
& ACS/SBC/F140LP & 1519.3 &  &  \multirow{2}{*}{2.60$\pm0.13$}  &  \multirow{2}{*}{0.06$\pm0.03$}  &  \multirow{2}{*}{--$^b$}  &  99$\pm$23  \\ 
& WFC3/UVIS/F225W & 2358.4 &  &    &    &   &  32$\pm$7  \\ 
\noalign{\smallskip}
 \hline \hline
    \end{tabular}}
\begin{minipage}{\linewidth}
     $\textbf{Notes.}$  Uncertainties at the 1$\sigma$ level. \\
     $^a$Extragalactic extinction on top of the foreground value, using \citet{cardelli_relationship_1989} extinction curve with $R_\mathrm{V} = 3.1$.\\
    $^b$The fit is over-constrained here.
\end{minipage}
\end{table*}

The exact conversion of count rate to fluxes (extinction-corrected or not) depends on the exact SED, which in turn is what we are trying to solve for. Therefore both the SED and the conversion factor need to be derived simultaneously. We do this again following \paper\ assuming a simple powerlaw for the SED. In all instances, we checked our assumption of a powerlaw SED looking that model - data residuals did not deviate strongly (3$\sigma$) from zero and that the $\chi^2$ values were reasonable, although we caution that the number of degrees of freedom is particularly uncertain when parameter ranges are constrained using priors \citep[see below and discussion in][]{andrae_dos_2010}. 

The extinction along the line of sight was modelled using the \citet{cardelli_relationship_1989} extinction curve with $R_\mathrm{V} = 3.1$ and {\ebv$_\mathrm{G}$ = 0.0747} \citep{schlafly_measuring_2011}. Another component with \ebv = 0.055$\pm$0.03 and redshift $z = 0.001568$ was added to model extinction from the host galaxy and local to the system, where the statistical\footnote{It is worth noting that in all likelihood the uncertainty on \ebv\ is larger owing to uncertainties related to the extinction curve or even differences between extinction curves, but these systematics are harder to quantify.} uncertainty on \ebv\ comes from \citet{grise_ultraluminous_2008} and is discussed in more detail in Section~\ref{sub:extinction}. The uncertainty introduced by the extinction in the final fluxes was taken into account by allowing \ebv\ to vary in tandem with the model parameters, but limiting its range of possible values by introducing a Gaussian prior on \ebv\ based on the measurement quoted above. 

 The fluxes derived from the modelling as well as constraints on the indexes of the powerlaws are reported in Table~\ref{tab:hst_fluxes}. In all instances we found \ebv\ consistent with the value quoted above and further constraints could not be achieved from the \hst\ broad filters.

The 2003 \hst\ observations probing the high state were well described by a single powerlaw (Table~\ref{tab:hst_fluxes}) with $\alpha = 3.4\pm0.1$. Considering the UV and optical/IR fluxes together, the 2016 observations from the low state were instead not well described by a single powerlaw ($\chi^2$/dof = 92/4), so we split them into two datasets. We found that fitting the IR/optical only (i.e. not considering the UV data) to a single powerlaw provided a reasonable fit ($\chi^2$/dof = 4.1/2) yielding $\alpha = 3.24\pm0.08$. Fitting the remaining UV data (the F225W and F140LP filters) with a single powerlaw we found $\alpha = 2.60\pm0.13$. Such $\alpha$ value is inconsistent with that derived from the  IR/optical data and thus supports the idea that these dataset do not follow a single powerlaw. Note the $\chi^2$ here loses its value as goodness of fit as the fit is over-constrained. 

We can see that the optical fluxes for both the low and high states are consistent within uncertainties. There is only evidence for variability in the near-IR (in the F814W filter). The constant optical variability seems, on the one hand, at odds with the variability observed in the B band with VLT observations by \citet{grise_ultraluminous_2008}. One reason could be the X-ray flux might have been lower during the \hst\ observations in the putative high state. Even if this is the case, the overall optical variability is small, a factor $\sim$1.25 at most as reported by \citet{grise_ultraluminous_2008}. Therefore uncertainties related to the exact X-ray flux level at the time of the \hst\ observation will not strongly impact our results. On the other hand, the non-variable optical emission and associated IR variability is in very good agreement with the work of \citet{sathyaprakash_multi-wavelength_2022}. Thus the apparent lack of optical variability with variable X-rays may be due to `luck' and the limited sampling here. For this reason we cannot completely exclude that the observed lack of optical variability is also consistent with \citet{grise_ultraluminous_2008}, as the authors showed the X-rays are not clearly correlated with the optical \citep[and similar conclusion may be drawn from][]{sathyaprakash_multi-wavelength_2022}. 
\section{Multi-band spectral fitting} \label{sec:multiband}

\subsection{Extinction}\label{sub:extinction}
The value of the extinction towards \theulx\ remains somewhat contested. \citet{liu_optical_2007} found discrepancy in the isochrones between two color-magnitude diagrams and argued these two diagrams could be reconciled assuming an extra $E(B-V) = 0.22$ mag on top the value along the line of sight (taken to be 0.11 mag in that work). \citet{grise_ultraluminous_2008} on the other hand showed that the discrepancy most likely arises due to a problem in the Padua HST/ACS isochrones, and that the discrepancy goes away if one compares instead Johnson-Cousins isochrones and assumes line-of-sight extinction only. \citet{grise_ultraluminous_2008} also reports \ebv\ = 0.13$\pm$0.03 mag from the Balmer decrement of an unpublished spectrum.

To measure the extinction towards \theulx, we extracted a spectrum using a 1.247\arcsec\ circle (matching the PSF FWHM) around the position of NGC~1313~X--2 in data cube 1. We corrected this spectrum from foreground extinction using \ebv = 0.0747 \citep{schlafly_measuring_2011} and the \citet{cardelli_relationship_1989} extinction curve with $R_\mathrm{V} = 3.1$. From this extinction-corrected integrated spectrum we measured $F$(\ha) = 4.4$\pm$0.2 and $F$(\hb) = 1.48$\pm$0.2 in units of 10$^{-16}$ \fluxcgs. The Balmer decrement is thus 2.97$\pm$0.45, which given the large uncertainties would be consistent with the case B recombination theoretical ratio \citep{osterbrock_astrophysics_2006} and argue against any additional extinction, in agreement with \cite{zhou_very_2022}. Assuming a typical theoretical ratio of 2.863 we find \ebv\ = 0.11$\pm$0.47 mag i.e. \ebv $<$ 0.18 mag for the total extinction. This conclusion does not change if we use the datacube prior to the sky residual correction. Therefore owing to the large uncertainties we adopt (and support) the value measured by \citet{grise_ultraluminous_2008} \citep[see also][]{zhou_very_2022}. 

Again it is useful to compare these estimates with those derived from the X-ray spectral fitting. Considering the relationship between the neutral hydrogen absorption column (\nh) and reddening found by \citet{guver_relation_2009} using simultaneous X-ray and optical observations of Galactic supernova remnants:
\begin{equation}
    N_\mathrm{H} = (2.21\pm0.09) \times 10^{21} A_\text{V}
\end{equation}
for \theulx\ would imply \citep[using the \ebv = 0.13$\pm0.03$ mag reported by][]{grise_ultraluminous_2008} $N_{\text{H}} = (9\pm2)\times10^{20}$cm$^{-2}$, which is in good agreement with the values derived from the broadband spectrum from the more plausible models (Section~\ref{sub:spectral_fitting}), supporting the low level of extinction. A more refined, subsequent work found significantly higher relationship \citep[$N_\mathrm{H} = (2.87\pm0.12) \times 10^{21} A_\text{V}$;][]{foight_probing_2016}. According to the authors, this relationship is also more appropriate when the ISM abundances from \citet{wilms_absorption_2000} are used in the derivation of \nh. With this relationship we find $N_{\text{H}} =12\pm3\times10^{20}$cm$^{-2}$, which is in even better agreement with the values derived from spectral fitting. However, we caution that since we are using subsolar abundances in our derivation of \nh\ below, this agreement is only orientative. 

Both the X-ray absorption and extinction are lower than found in NGC~1313~X--1, which again supports both the higher extinction found in NGC~1313~X--1 \citep{gurpide_quasi_2024} and the lower extinction found here. This is also consistent with the position of \theulx, located far away from the nucleus compared to NGC~1313~X--1, which is instead embedded in the arms of the galaxy where additional dust might be expected.

\subsection{Spectral fitting} \label{sub:spectral_fitting}
We used \xspec\ version 12.13.1 to perform spectral fitting of the broadband data. We modelled neutral absorption along the line of sight with two \tbabs\ components \citep{wilms_absorption_2000} with cross-sections from \citet{verner_atomic_1996}. The first component was frozen to the value along the line of sight \citep[6.29$\times10^{20}$ cm$^{-2}$;][]{hi4pi_collaboration_hi4pi_2016} to model the contribution from the Galactic absorption. The second component models any local absorption to \theulx\ and was modelled using the more flexible \texttt{tbfeo} model component with the redshift set to that of NGC~1313 ($z = 0.001568$). Although the metallicity around NGC~1313~X--2 remains somewhat uncertain \citep[see discussion in][]{ripamonti_metallicity_2011}, we set the abundances of oxygen and iron to 50\% of the Solar value\footnote{Technically this is 50\% of the ISM abundances set by \citet{wilms_absorption_2000}, which are confusingly called `solar' in \xspec. We also make no distinction between the two, as more level of refinement is unnecessary for the data at hand.} in an attempt to account for the subsolar environment of the source/galaxy. We note that it is likely that more level of refinement is an overkill considering the medium-resolution X-ray data we are modelling. We discarded the MOS cameras from this part of the analysis so as to have similar weighting over all bands in the fit, but we note similar results are obtained if we include these datasets.

Following \citet{gurpide_quasi_2024}, we considered the same physically motivated models (namely the \texttt{diskir}; \citet{gierlinski_reprocessing_2009} and the \texttt{sirf}; \citet{abolmasov_optically_2009}) and a third, phenomenological model (\texttt{phenomenological} hereafter). 

The \texttt{diskir} can be considered an extension of the classical \texttt{diskbb} \citep{mitsuda_energy_1984}, including Comptonization and self-irradiation effects, the latter important in describing the UV/optical bands. Implicitly, this model assumes a sub-Eddington, optically thick, geometrically thin accretion disk. Instead, the \texttt{sirf} can be considered an extension of this model to the super-Eddington regime, where the wind/cone alters the pattern of self-irradiation of the flow. However, in this model the outer parts of the accretion flow, where the disk remains sub-Eddington, are absent. With the \texttt{phenomenological} model, we attempt to describe or account for the accretion flow around the NS in NGC~1313~X--2. This model is mainly based on the work of \citet{walton_evidence_2018} and consists of two dual thermal components \citep[\texttt{diskbb};][]{mitsuda_energy_1984} and a \texttt{cutoffpl} modelling the emission from the accretion column at higher energies.

Along with the continuum with added a Gaussian line around $\sim$1 keV as the spectra presented the usual prominent soft residuals \citep{middleton_diagnosing_2015, gurpide_long-term_2021} attributed to highly-blueshifted super-Eddington winds \citep{pinto_resolved_2016, kosec_searching_2018}. Such component was strongly required by the fit, since, for instance, for the \texttt{diskir} it provided a $\Delta \chi^2 > 30$ improvement for both the high and low states for 3 additional degrees of freedom. We further noted that its parameters remained constant within the two states, therefore we tied its parameters between them to reduce parameter degeneracies and degrees of freedom.

\begin{figure}
    \centering
    \includegraphics[width=0.49\textwidth]{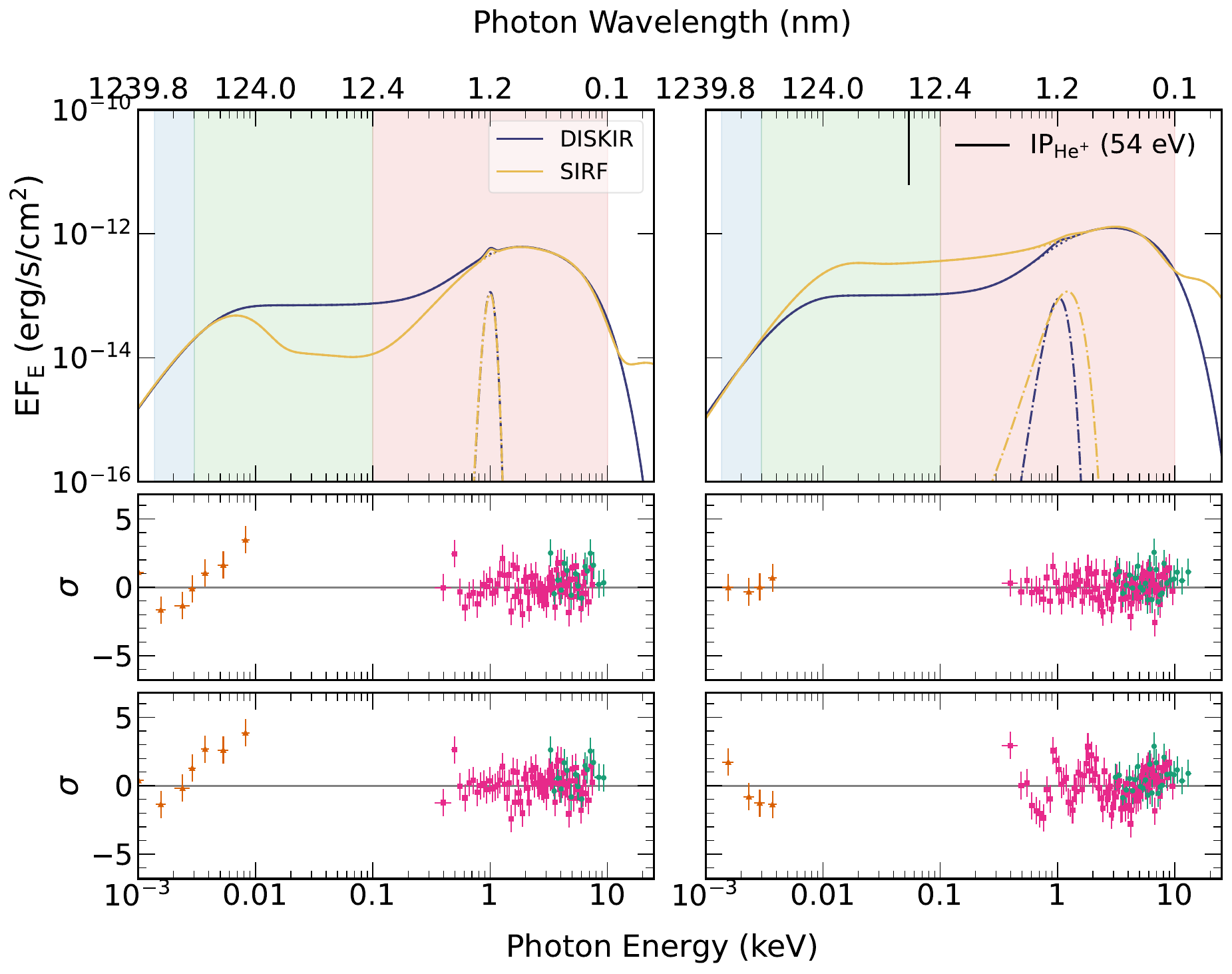}
    \caption{Broadband best-fit models and residuals. (Top panels) Spectra of the low (left) and high (right) states for the \texttt{diskir} and \texttt{sirf} spectral models (solid lines as labelled). The dash-dotted lines show the Gaussian emission line at $\sim$1 keV. A black vertical tick indicates the ionization potential of He$^{+}$. (Middle) Residuals for the \texttt{diskir} model, rebinned to a minimum significance of $3\sigma$. Orange, pink and green markers show data from \hst, EPIC-pn and \nustar-FPMA detectors, respectively. (Bottom) As per middle panel but for the \texttt{sirf} model.}
    \label{fig:broadband_seds}
\end{figure}

We started by testing whether the broadband data could be described without the inclusion of any additional component. We excluded the \texttt{phenomenological} from these initial tests as it makes no account of the optical data (something we already showed in \paper\ for NGC~1313~X--1). Figure~\ref{fig:broadband_seds} shows the best-fit \texttt{diskir} and \texttt{sirf} models and residuals. None of the models is able to describe accurately the broadband SED. This is particularly noticeable in the residuals of the low states, which clearly indicate a deficiency in the model, particularly due to the more constraining UV data. We therefore considered the addition of a \texttt{bbody} component contributing to the optical data. 

In general we found the models with the \texttt{bbody} to have too much flexibility with respect to the data quality or coverage. This lead to several local minima and many parameters being degenerate (for instance \nh\ with the soft component or  \texttt{bbody} parameters). Therefore we restricted the flexibility of the models as follows.

For the \texttt{phenomenological} model, faced with a lack of constraints on the shape of the pulsed component, it is customary \citep{robba_broadband_2021, amato_ultraluminous_2023} to fix the energy cutoff \ecutoff\ and $\Gamma$ of the \texttt{cutoffpl} to the average values derived for pulsating ULXs \citep[$\Gamma = 0.59$; \ecutoff = 7.9 keV;][]{walton_evidence_2018}. Here too we found that despite the additional coverage provided by \nustar, both \ecutoff\ and $\Gamma$ could not be constrained simultaneously. We therefore decided to fix \ecutoff\ = 7.9 keV while leaving $\Gamma$ free to vary, as the latter parameter should have a stronger impact on the extrapolation to low energies. For the high state we found the normalization of the \texttt{cutoffpl} tended to zero. While this may call into question the need for this component, we opted to fix $\Gamma = 0.59$ for consistency with the modelling of the low state. For the \texttt{sirf} we noted $\dot{m}_\text{eje}$\ could not be constrained but a lower value was preferred by the fit, so we fix it to 10. Similarly, we set $kT_\mathrm{e} = 30$ keV for the \texttt{diskir} as this parameter mainly affects the high-energy tail. We further tied the \texttt{bbody} parameters between the high and low states given the little observed optical variability. This also helped `force' the \texttt{bbody} into the optical/UV bands.

Lastly, we decided to also tie \nh\ between the high and low states after some initial tests. In general, it was possible to find solutions were both states had consistent level of \nh, but for the \texttt{diskir} and \texttt{sirf} -- which are also the more convoluted and therefore degenerate models -- there were other minima (with nearly equivalent $\chi^2$) preferred by the minimization routine, where this was not the case. For the \texttt{phenomenological} best-fits, we found both states had comparable \nh\ ($N_\mathrm{H}^\mathrm{high} = 11.7_{-0.8}^{+0.9} \times10^{20}$ cm$^{-2}$ and $N_\mathrm{H}^\mathrm{low} = 11\pm1 \times 10^{20}$ cm$^{-2}$). Such apparent lack of \nh\ variability is in very good agreement with our previous analysis \citep{gurpide_long-term_2021}. A firm conclusion about the variability of \nh\ cannot be drawn without physically motivated models, but in our view the data at hand suggest variations of \nh\ are most likely driven by parameter degeneracies. We therefore considered tying \nh\ another reasonable way to limit parameter degeneracies. 

\begin{figure*}
    \centering
    \includegraphics[width=0.98\textwidth]{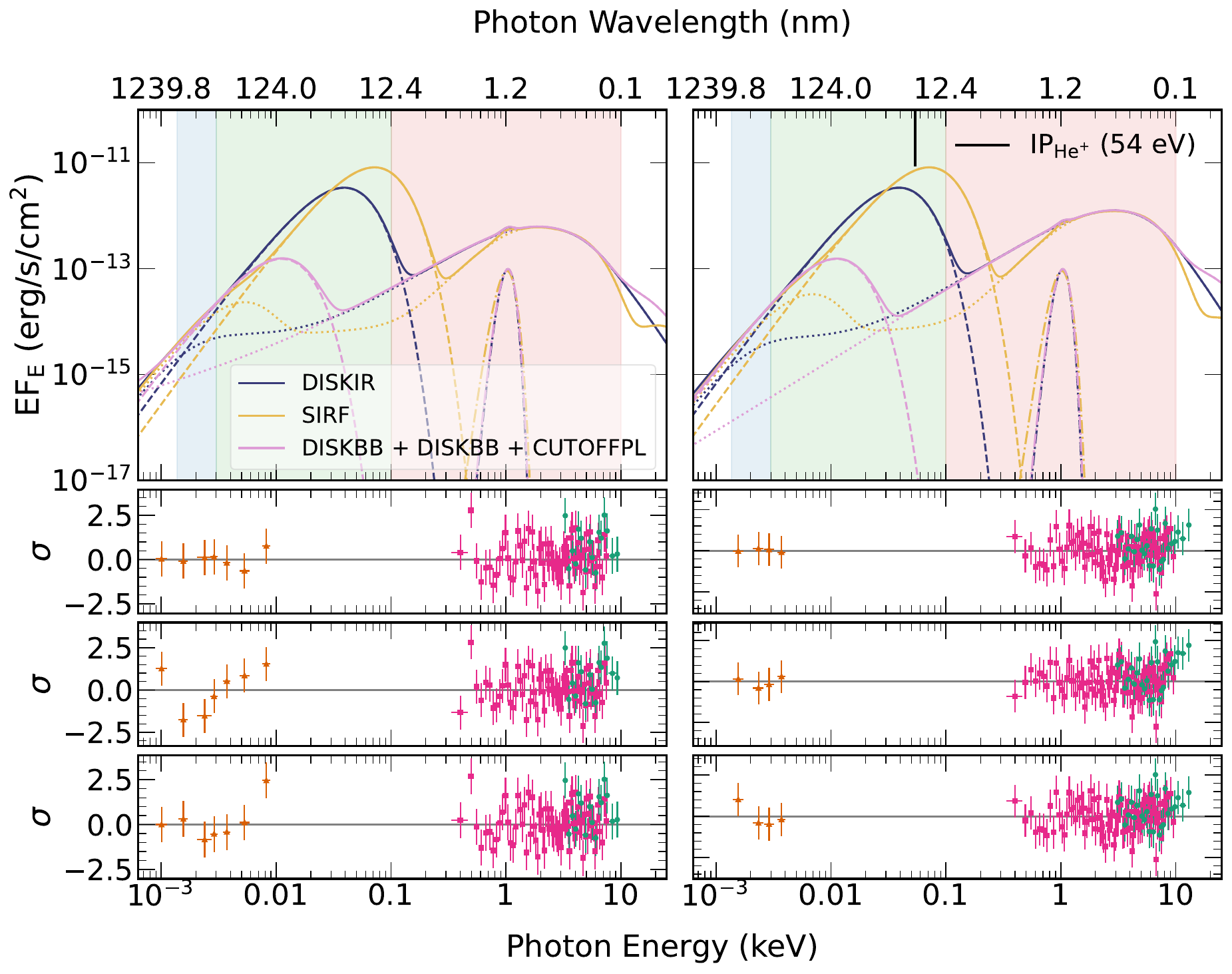}
    \caption{As per Figure~\ref{fig:broadband_seds} but now including a \texttt{bbody} emission (dashed lines) contributing to the optical/UV emission and including the \texttt{phenomenological model}. (Top panels) Spectra of the low (left) and high (right) states for the \texttt{diskir}, \texttt{sirf} and \texttt{phenomenological} spectral models (solid lines). The main contribution from the accretion flow is shown as a dotted line. (Bottom panels) Residuals for the \texttt{diskir} (top), \texttt{sirf} (middle) and \texttt{phenomenological} (bottom) models. Note the change in scale of the y axes with respect to Figure~\ref{fig:broadband_seds}.}
    \label{fig:broadband_seds_bbody}
\end{figure*}

Figure~\ref{fig:broadband_seds_bbody} show the best-fit spectral models and residuals while Table~\ref{tab:sed_modelling} show the resulting best-fit parameters and fit statistics. De-absorbed fluxes in different bands are also given and were computed with the pseudo-model \texttt{cflux}.

All models provide statistically acceptable fits to the data, with the \texttt{diskir} providing marginally better fits. This is because the \texttt{phenomenological} and \texttt{sirf} have difficulties in fitting the UV data but not to the point of being statistically rejected. Better fits could be easily accomplished relaxing some of our fitting constraints. Therefore based on the data alone we cannot reject any of the models.

We note that these best-fit models presented here are not the only possible solution. For instance, we have found it is also possible to explain the data under the \texttt{sirf}, with a very cold \texttt{bbody} fitting the F814W fluxes with little contribution in the optical bands and an overall lower UV flux. However, in fitting these data our goal is not so much aimed at finding the best physical model describing the accretion flow in \theulx, but rather in obtaining realistic extrapolations to the UV supported by the data, which we can now test against the nebular emission.  

\begin{table*}
    \begin{center}
     \caption{Best-fit parameters to the multi-band SED of \theulx.}
    \label{tab:sed_modelling}
    \begin{tabular}{cccc} 
    \hline
    \noalign{\smallskip}
      Parameter  &   Units   & Low State                & High State   \\
      \noalign{\smallskip}
    \hline
    \noalign{\smallskip}
                 &           &  \multicolumn{2}{c}{\texttt{DISKBB} + \texttt{DISKBB} + \texttt{CUTOFFPL} + \texttt{BBODY}} \\
\noalign{\smallskip}
\hline
\noalign{\smallskip}
\nh$^a$             &10$^{20}$cm$^{-2}$&  \multicolumn{2}{c}{11.4$\pm0.6$}  \\ 
 $T_\text{in}$   & keV              &  0.52$\pm0.06$              &  0.8$\pm0.1$ \\
$T_\text{in}$    &  keV             &  1.1$\pm0.1$        &  1.5$\pm0.1$  \\
$\Gamma$         &                  &  1.38$_{-0.05}^{+0.13}$     &  0.59      \\
\ecutoff         & keV              &  7.9                        &   7.9        \\
$T$               & K                    &       \multicolumn{2}{c}{$33653_{-4641}^{+6962}$}   \\
$R$               & $R_{\sun}$            &    \multicolumn{2}{c}{$10_{-3}^{+5}$}        \\
$\chi^2/$dof     &	                &  \multicolumn{2}{c}{274.6/292}         \\
$p$-value        &                  &  \multicolumn{2}{c}{0.77}              \\
$L_\mathrm{opt}^b$  & 10$^{37}$\,\lumcgs\   & $1.71\pm0.06$       &  $1.61\pm0.06$             \\ 
$L_\mathrm{UV}^c$   & 10$^{39}$\,\lumcgs\   & $1.1^{+1.2}_{-0.5}$ &  $0.6\pm0.3$                \\
$L_\mathrm{X}^d$    & 10$^{39}$\,\lumcgs\   & $3.2\pm0.03$       &  $5.94\pm0.04$               \\
$L_\mathrm{bol}^e$  & 10$^{39}$\,\lumcgs\   & $4.4^{+1.4}_{-0.5}$ &  $7^{+3}_{-0.7}$              \\
  \noalign{\smallskip}
  \hline
  \noalign{\smallskip}
 &       &              \multicolumn{2}{c}{\texttt{DISKIR} + \texttt{BBODY}}\\
  \noalign{\smallskip}
  \hline
    \noalign{\smallskip}
\nh\                    & 10$^{20}$cm$^{-2}$ &      \multicolumn{2}{c}{11.4$\pm0.6$}           \\
$kT_\text{disk}$	    &	keV	             &      0.53$\pm0.06$      & 0.8$\pm1$    \\
$\Gamma$	            &		             &	$>$4.1                & 4.6$_{-0.5}^{+0.8}$    \\ 
$kT_\text{e}$           &	keV	             &	 \multicolumn{2}{c}{30}            \\
$L_\text{C}/ L_\text{D}$&                    &  0.7$\pm0.2$                        & 0.7$\pm0.3$                        \\
$R_\text{irr}$          &	\rin\            &  1.02$\pm$0.01               & 1.05$_{-0.02}^{+0.02}$               \\
$f_\text{out}$          &	10$^{-3}$        &  1.9$_{-0.9}^{+0.7}$                     & 0.9$_{-0.5}^{+0.6}$                \\
log(\rout)	            & \rin\              &	4.8$\pm0.1$         & $>4.7$                       \\
$T$               & K                    &       \multicolumn{2}{c}{$>58023$}   \\
$R$               & $R_{\sun}$            &    \multicolumn{2}{c}{$3_{-4}^{+7}$}        \\
$\chi^2/$dof            &                    &   \multicolumn{2}{c}{267.6/289}                       \\
$p$-value               &                    &   \multicolumn{2}{c}{0.8}                   \\
$L_\mathrm{opt}$    & 10$^{37}$ \lumcgs\   &  $1.61\pm0.07$         &  $1.5^{+0.1}_{-0.04}$       \\ 
$L_\mathrm{UV}$     & 10$^{39}$\,\lumcgs\  &  $10^{+1}_{-9}$        &  $10^{+2}_{-9}$            \\
$L_\mathrm{X}$      & 10$^{39}$\,\lumcgs\  &  $3.30^{+0.05}_{-0.1}$ &  $6.06^{+0.04}_{-0.1}$   \\
$L_\mathrm{bol}$    & 10$^{39}$\,\lumcgs\  &  $13^{+1}_{-8}$        &  $16^{+2}_{-8}$           \\
\noalign{\smallskip}
\hline
\noalign{\smallskip}
&   &  \multicolumn{2}{c}{\texttt{SIRF}+ \texttt{BBODY}}\\
\noalign{\smallskip}
\hline
\noalign{\smallskip}
\nh\ & 10$^{20}$cm$^{-2}$               & \multicolumn{2}{c}{2.3$\pm0.8$}      \\
$T_\text{in}$	  & keV                  & 0.38$\pm0.02$              &  0.43$\pm0.02$ \\
\rin\             & 10$^{-3}$ \rsph\     &  0.320$\pm0.002$           &  0.350$\pm0.002$  \\
\rout\            & $R_\text{sph}$       & 190$_{-14}^{+13}$          &  149$_{-13}^{+14}$ \\
$\theta_\text{f}$ & \degr             & 16.7$\pm0.2$                    &  15.9$\pm0.2$ \\ 
\mout             & $\dot{M}_\text{Edd}$ &  \multicolumn{2}{c}{10} \\
$T$               & K                    &       \multicolumn{2}{c}{$>116045$}   \\
$R$               & $R_{\sun}$            &    \multicolumn{2}{c}{$1_{-1}^{+3}$}        \\
$\chi^2/$dof            &                    &   \multicolumn{2}{c}{280.8/293}                       \\
$p$-value               &                    &   \multicolumn{2}{c}{0.69}                   \\
$L_\mathrm{opt}$& 10$^{37}$ \lumcgs\  &   $1.75\pm0.04$       &  $1.59\pm0.06$       \\
$L_\mathrm{UV}$ & 10$^{39}$ \lumcgs\  &  $19.6^{+9}_{-11}$  &  $20^{+10}_{-15}$         \\
$L_\mathrm{X}$  & 10$^{39}$ \lumcgs\  &  $7^{+7}_{-4}$         &  $10^{+7}_{-4}$       \\
$L_\mathrm{bol}$& 10$^{39}$ \lumcgs\  &  $27^{+16}_{-18}$         & $29^{+17}_{-19}$  \\
  \noalign{\smallskip}
\hline
\hline
\end{tabular}
\begin{minipage}{\linewidth}
\textbf{Notes.} Uncertainties are given at the 1$\sigma$ level. Parameters without error bars were fixed at the given value. All luminosities are corrected for absorption.\\
For the \texttt{sirf} model we adopted a velocity law exponent of the wind of $-$0.5 (i.e. parabolic velocity law), adiabatic index $\gamma$= 4/3 (appropriate for radiation-pressure supported gas) and $i = 0.5^\circ$ \citep[i.e. an edge-on system; e.g.][]{sathyaprakash_discovery_2019}. \\
$^a$Extra-galactic absorption column. The galactic component was frozen to the value along the line of sight \citep[6.29$\times10^{20}$ cm$^{-2}$;][]{hi4pi_collaboration_hi4pi_2016}.\\
$^b$Optical luminosity, defined in the 4000--9000\AA\ band.\\
$^c$UV luminosity (3\,eV, 4000\AA\ -- 0.1\,keV).\\
$^d$X-ray luminosity (0.1--10\,keV).\\
$^e$Bolometric luminosity (10$^{-4}-10^{3}$\,keV).\\
\end{minipage}
\end{center}
\end{table*}

\section{Constraints on the EUV emission of \theulx} \label{sec:cloudy}
As apparent from our multi-band modelling (Section~\ref{sec:multiband}), there is a wide range of possible UV/EUV extrapolations compatible with the data, highlighting the uncertainty in extrapolating the ULX SED to the UV. Each of the models presented in Section~\ref{sub:spectral_fitting} inherently makes a prediction about the UV/EUV emission in \theulx, which we can test against the observed nebula. In particular, the model including the \texttt{sirf} predicts a powerful UV source ($L_\mathrm{UV} \gtrsim 10^{40}$ erg/s), while the \texttt{diskir} predicts $L_\text{UV} > 10^{39}$ erg/s. The \texttt{phenomenological} predicts the lowest UV flux, at $\sim 1 \times 10^{39}$ erg/s. Here we take advantage of the EUV-sensitive \heii\ line \citep{pakull_detection_1986, pakull_optical_2002} to assess which model predicts a UV flux compatible with our upper limit derived in Section~\ref{sub:heii}.

To this end, we use \cloudy\ v.22.02 \citep{ferland_2017_2017} to produce a spherically symmetric cloud -- mimicking \theulx\ and its surrounding nebula -- ionized by the (deabsorbed) SEDs presented in Section~\ref{sec:multiband}. We then integrate the \heii\ emission produced by this cloud and compare it with the upper limit derived in Section~\ref{sub:heii}. Since the \heii\ line requires shock velocities \vs\ $\sim$275 km/s to be collisionally excited \citep[e.g. as observed in the microquasar S26;][]{pakull_300-parsec-long_2010}, we can assume \heii\ is produced purely by photo-ionization in the \theulx\ bubble, given that \vs\ $\sim$80 km/s \citep{pakull_optical_2002, ramsey_optical_2006, zhou_very_2022}. 

\begin{figure}
    \centering
    \includegraphics[width=0.49\textwidth]{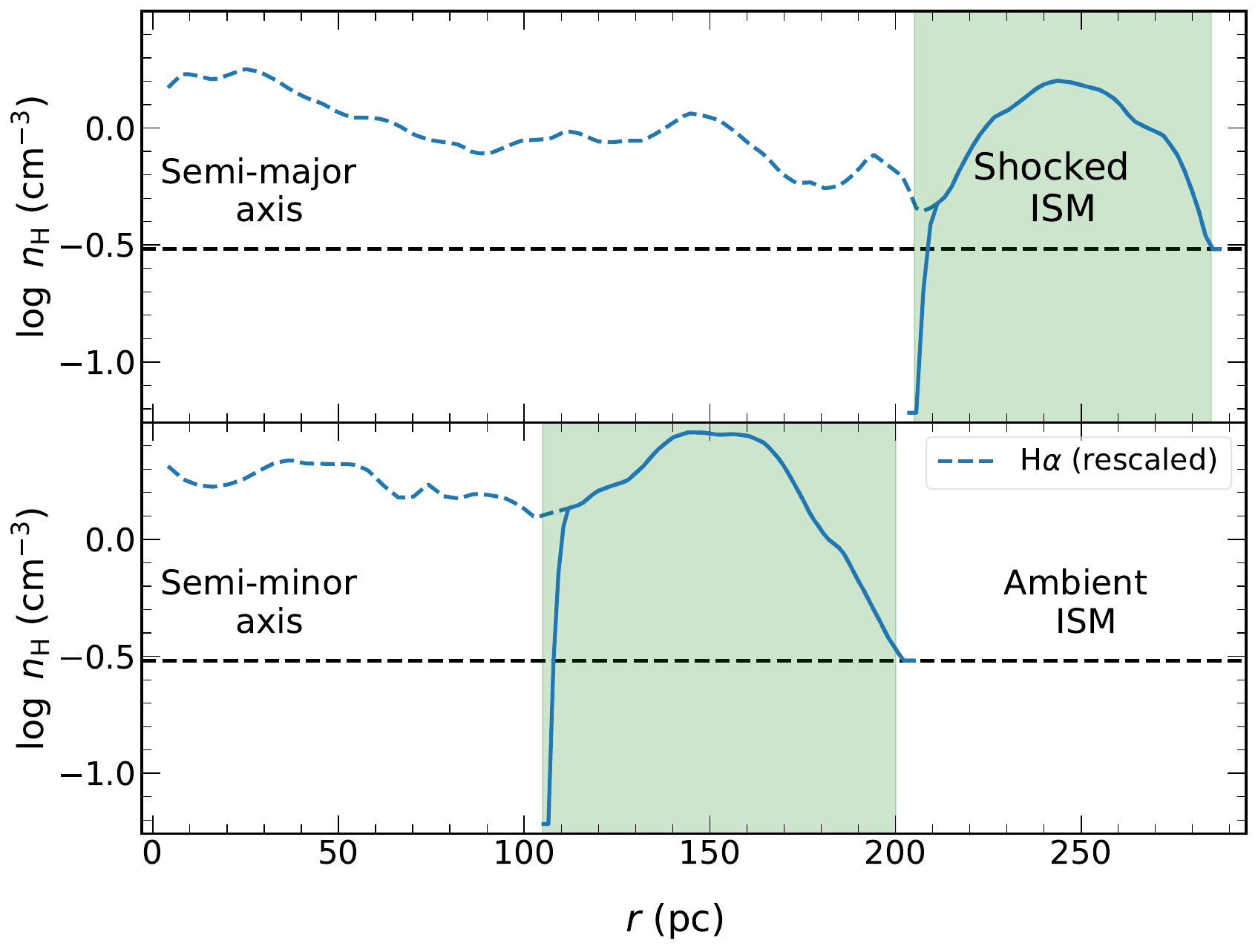}
    \caption{Hydrogen number density profiles (solid lines) for the bubble around \theulx\ assumed in our photo-ionized modelling. The dashed lines show the (re-scaled) H$\alpha$ profiles along the semi-major (top) and semi-minor axes (bottom) of the bubble used to derive $n_\mathrm{H}(r)$. The H$\alpha$ fluxes have been re-scaled to number density by matching the flux outside the bubble to the ambient ISM density. In the regions interior to the shocked ISM (green area) we assume the density is effectively zero. The regions used to extract the H$\alpha$ profiles are indicated in Figure~\ref{fig:bubble_ha}. }
    \label{fig:density_law}
\end{figure}

To assume a realistic density profile, we considered the effects of the winds emanating from \theulx\ in carving and shocking the ISM \citep{weaver_interstellar_1977, siwek_optical_2017}. Accordingly, we should expect a thin shell of dense ($n >$ \nism) swept-up ISM in the outer edge of the bubble. Figure~\ref{fig:density_law} shows two profiles in H$\alpha$ along the semi-major and semi-minor axes of the bubble. These were extracted using the regions shown as blue boxes in Figure~\ref{fig:bubble_ha}. We identify the outer luminosity rise and decay seen at the edges of the bubble (green areas in Figure~\ref{fig:density_law}) as this thin shell of swept-up ISM. Theoretically, H$\alpha$ should decrease in brightness towards the center due to the sharp subsequent decrease in density \citep{siwek_optical_2017}, but in practice projection effects prevent us from observing this. Therefore in the inner regions to the shocked ISM, we assume the bubble is hollow, as the density in these regions is too low for any meaningful impact in our results \citep{siwek_optical_2017}. Therefore based on the H$\alpha$ profiles (Figure~\ref{fig:density_law}), we assumed $R_\mathrm{in} = 105, 205$ pc and $R_\mathrm{out} = 200, 285$ pc for the semi-minor and semi-major axes. 

Next we note \cite{zhou_very_2022} estimated the density of the undisturbed ISM to be \nism = 0.45$\pm$0.02 cm$^{-3}$ based on scaling relationships established for radiative shocks \citep{dopita_spectral_1996}, which link the luminosity of some diagnostic line (e.g. \hb) with the shock velocity and the density of the ISM (their Equation 3.4). In passing, we note that this scaling may not be `universal' as it is derived for a certain set of abundances as described in \citet{dopita_spectral_1996}. How or whether this scaling varies with abundances has not been studied to the best of our knowledge. Fitting the \hb\ line from the integrated nebular spectrum (Figure~\ref{fig:zap_bubble}) with a Gaussian and correcting for foreground extinction we find very good agreement with the \nism\ estimate derived by \citet{zhou_very_2022}. However, such estimate neglects that a fraction of \hb\ is produced by photo-ionization of the ULX, which must be discounted. 

The bubble nebula shows broadened lines and line ratios characterised of supernova remnants, all clear signatures of being shock-ionized \citep{pakull_optical_2002, zhou_very_2022}. However, \citet{zhou_very_2022} showed clearly that there are two bright patches inside the bubble west of the ULX (clearly seen in Figure~\ref{fig:bubble_ha}) where the ratios are also affected by photo-ionization (see also G\'urpide et al. in prep.). We determined that about 40\% of the \hb\ luminosity is produced in this region. We then need to estimate what fraction of this \hb\ luminosity comes from photo-ionzation. We can get a rough idea by measuring the \hb\ surface brightness outside this region in the other areas of the bubble where we know \hb\ is excited by shocks. With this simplistic approach we determine that about 65\% of the \hb\ luminosity in the bright patch is due to photo-ionization, hence we determine that $\approx$25\% of the \textit{total} \hb\ luminosity in the bubble is due to photo-ionization. Hence we revisit the estimate on $n_\mathrm{ISM}$ downwards by 25\% to 0.34 cm$^{-3}$\footnote{Note that \cloudy\ takes hydrogen density ($n_\mathrm{H}$). Therefore we take 90\% of this value as our final estimate for $n_\mathrm{H}$.}. 

We then re-scaled the H$\alpha$ profiles so that the flux at the outer edge matched this estimated \nism\ density. Using this rescaling we find that the density within these shocked ISM region reaches about 5 times higher than \nism. From the numerical calculations of \citet{siwek_optical_2017}, we can see that for a wind power $P_\mathrm{w} \approx 10^{40}$ erg/s, the thin shell of swept-material can reach densities $\approx$ 100\nism. Given the factor 10 lower mechanical power in \theulx, we find the densities estimated in this manner quite reasonable. Note also that by re-scaling the H$\alpha$ to a density number we naturally find the semi-minor axis to have a higher density. This would be consistent with our physical intuition that the oblate appearance of the bubble is due to density differences along the different axis. We further note that the electron-density sensitive [S {\sc ii}]$\lambda$6716/[S {\sc ii}]$\lambda$6731 ratio is at the limit of the low-density regime (from the integrated spectrum, we find [S {\sc ii}]$\lambda$6716/[S {\sc ii}]$\lambda$6731 $\approx$1.43$\pm$0.02, consistent with \citet{zhou_very_2022}), indicating the density in the nebula must be $<$ 10 cm$^{-3}$.

We have also attempted to take into account the effects of shocks in heating the gas. We have done so by assuming the gas is in collisional equilibrium within \cloudy. In reality the gas is not in equilibrium as it is cooling effectively but we assume this is a good first order approximation. We adopted a temperature of the gas of {20,000--25,000 K}, typical of shock-excited bubbles \citep[e.g.][]{pakull_300-parsec-long_2010} but found the final line fluxes to be weakly affected even for the most extreme case (cf. Figures~\ref{fig:heII_cloudy_hot} and Figure~\ref{fig:heII_cloudy} below). Therefore we focus on the results for the default assumption of neutral ISM (resulting calculations for $T = $25,000 K can be found in the Appendix in Figure~\ref{fig:heII_cloudy_hot}).

  
There are two more quantities we need to consider with regards to the geometry: the filling factor $f_\mathrm{f}$ and the covering factor $\Omega$. The former assumes the gas is not uniformly distributed inside the thin shells, but rather it is formed of clumps with vacuum or empty space in between them, which obviously do not contribute to the opacity/emission. Note here the filling factor refers to the inside of the thin shells of the bubble, as the inner radii already take into account the likely depletion of gas inside of the bubble. The covering factor refers to the fraction of the bubble nebula that is actually covering the source and we consider it to approximate the fact that the bubble is a patchy ellipsoid (Figure~\ref{fig:bubble_ha}), as opposed to a complete structure. For the filling factor we have explored the sensitivity of our results to a wide range, namely $f_\mathrm{f}$ = 0.3, 0.5, 0.9. For the covering factor, we can obtain a sensible estimate by first measuring the area of the two hollow-looking patches east of the ULX and another darker area north of it. The ratio of the combined area of these `empty' patches to the total bubble surface is found to be about 0.85 and we take this as our estimate for $\Omega$. The predicted \heii\ fluxes decrease linearly with $\Omega$ e.g $\Omega$ = 0.85 reduces the predicted fluxes by 15\% compared to the case where $\Omega$ = 1. Thus, uncertainties introduced by changes in $\Omega$ are relatively small. 

Finally we need to set the abundances of the cloud. \citet{ripamonti_metallicity_2011} provides different estimates for the metallicity content of the bubble surrounding \theulx, based on empirical scalings or photo-ionization modelling. Given that in NGC~1313~X--1 \citet{gurpide_quasi_2024} determined $Z \approx 0.002$ (or \met\ = 7.9) and that the metallicity is observed to decrease with distance from the galactic nucleus in NGC~1313, we favour metallicities towards the lower end of the values estimated by \citet{ripamonti_metallicity_2011}, although we caution these are uncertain as their calculations neglected shocks and assumed constant number density. We thus ran our simulations for $Z = 0.003, 0.001$\footnote{In practice these were rescaled from the solar abundances from \citet{grevesse_chemical_2010} where $Z_{\sun}$ = 0.0134.} corresponding to \met = 8.0, 7.6, respectively, noting the metallicity does not strongly impact our results.

\begin{figure*}
    \centering
    \includegraphics[width=0.95\textwidth]{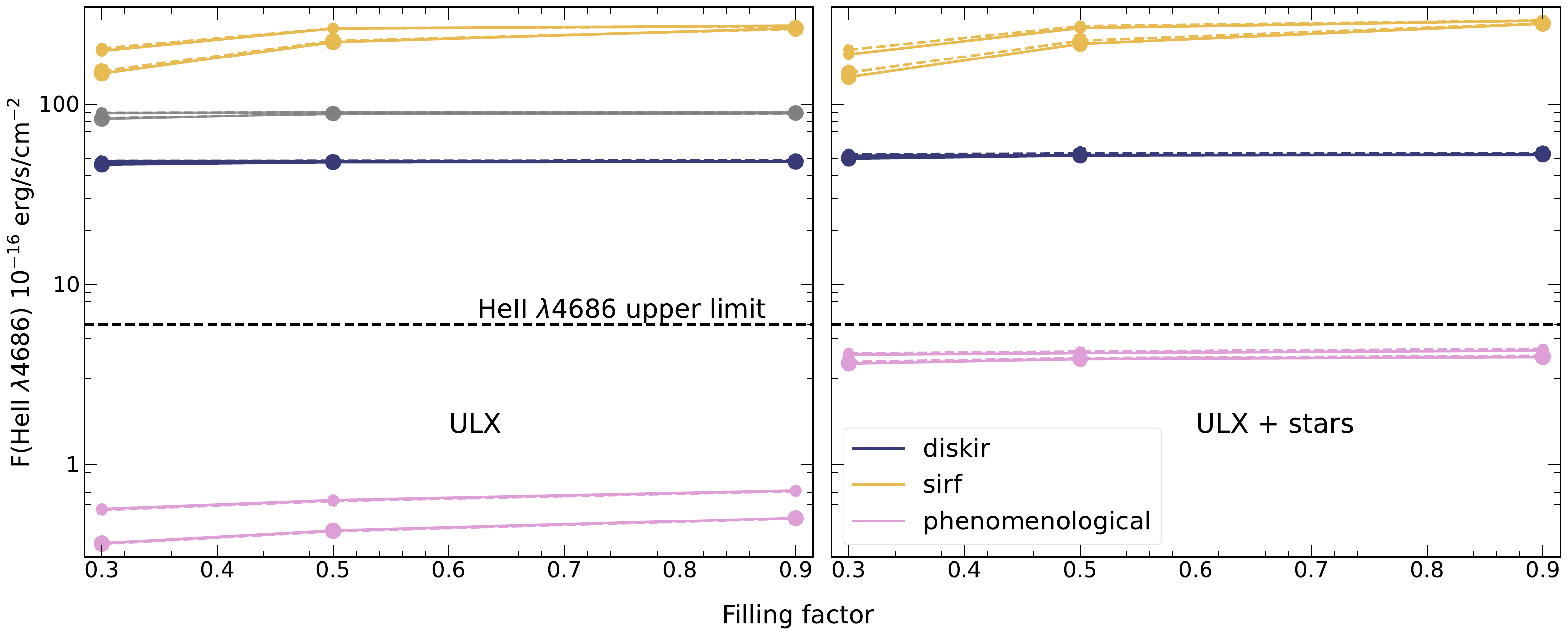}
    \caption{Predicted \heii\ flux for the three SEDs (high state) presented in Section~\ref{sec:multiband} when the ULX is the only source of ionization (left) and when the background stars are included (right). The dashed black line shows the extinction-corrected upper limit on \heii\ from the integrated spectrum of the whole bubble (Section~\ref{sub:heii}). Models that predict an \heii\ flux above the dashed line are too bright and therefore excluded. Results for the same model/$R_\mathrm{in}$/$Z$ are connected by lines. Thin and thick markers show the results for $R_\mathrm{in} = 105$ (semi-minor axis) and 205 pc (semi-major axis), respectively, while results for $Z = 0.001, 0.003$ are shown by solid and dashed lines, respectively. The gray lines are the \heii\ fluxes predicted using the \texttt{blackbody} inferred by \citet{abolmasov_optical_2008} in NGC~6946~X--1.} 
    \label{fig:heII_cloudy}
\end{figure*}

Figure~\ref{fig:heII_cloudy} (left panels) shows the predicted \heii\ fluxes based on the three different SEDs presented in Section~\ref{sec:multiband} for the high state. The results for the low state are shown in the Appendix (Figure~\ref{fig:heII_cloudy_appendix}), but because the differences between the high and low states were insignificant, we focus on the results for the high state. Similar results between the high and low states are expected as the main difference between the two states is in the X-ray band, where as alluded in the Introduction He$^+$ is no longer sensitive to. This simple results shows the \heii\ line cannot be used to probe the collimation of the hard X-ray emission, in agreement with the cautionary argument already made by \citet{abolmasov_optical_2007} \citep[for more details on this aspect we refer the reader to][]{beuchert_exploring_2024}.

We can see that base on the implied UV/EUV level by the \texttt{diskir} and \texttt{sirf} models, the \heii\ should have been detected with luminosity one or two orders of magnitude above our upper limit. Instead, the \texttt{phenomenological} predicts a \heii\ luminosity consistent with the derived upper limit. Differences due to the filling factor or metallicity are negligible. Also the smaller size of the semi-minor axis is offset by the implied higher density, with the resulting \heii\ being comparable to that derived using the semi-major axis morphology instead. 

From our photo-ionization models we find hydrogen column densities \nh\ $\approx$ (0.75--4.6) $\times 10^{20}$ cm$^{-2}$ which are roughly of the order of the extragalactic \nh\ derived from X-ray spectral fitting. This indicates our density estimate for the cloud is reasonable. There could be additional absorbing material closer to the ULX or from the host galaxy itself. 

There is inevitably some uncertainty with regards to $R_\mathrm{in}$ owing to projection effects and the PSF size (FWHM $\approx$25 pc at 4.25 Mpc), although the bubble edges are broader than the PSF, indicating they are resolved. Nevertheless, we have tested the sensitivity of our calculations to this uncertainty by halving the depth of the bubble edges by increasing $R_\mathrm{in}$ (i.e. setting $R_\mathrm{in} = 150, 242$ pc for the semi-minor and semi-major axes, respectively). We have found the predicted \heii\ luminosities remained roughly unaffected for the \texttt{diskir} and \texttt{phenomenological} models (changes of less than 10\%), while they dropped by $\approx$13, 30 and 35\% for the \texttt{sirf} model for filling factors $f_\mathrm{f} = 0.3, 0.5$ and 0.9 respectively. The predicted \heii\ luminosity does not vary for the former models because there is enough gas for most UV photons to be absorbed even assuming a thinner shell, in other words, the nebula is ionization bounded. For the \texttt{sirf}, the predicted \heii\ decreases because a fraction of the UV photons absorbed in the thicker models now escape the nebula. Variations in the predicted \heii\ luminosities are again small and indicate our conclusions are unchanged for any reasonable range of cloud depths. 

Similarly, we consider the dependency of our results to the assumed densities. The fact that the \heii\ predictions show little variability with the filling factor already indicates that all He$^+$ ionizing photons are being absorbed. Therefore, even if the density was higher than assumed it would not affect the predicted \heii\ intensities, because all ionizing photons have already been exhausted. We have explicitly verified this by running the same calculations but doubling the density at every radius in the density profile. The little dependency of \heii\ with the filling factor also indicates that even if the density was lower than assumed, the results would not be strongly affected. We tested this by running the calculations with the same density profile, but halving the density at every radius of the profile. As for the case above where we reduced the radius of the shell, we find the only affected model is the \texttt{sirf}, whose predicted \heii\ becomes comparable to that of the \texttt{diskir} for the lowest filling factors ($f_\mathrm{f} = 0.3, 0.5$). If we further halve the inner radius as above, for the lowest $f_\mathrm{f}$ the \texttt{diskir} and \texttt{sirf} still predict \heii\ fluxes more than 3 to 9 times  our derived upper limit. Therefore, unless we are dramatically overestimating the density of the gas, our results do not depend strongly on the assumed density.

Finally we have tested the effects of adding an additional source of ionising photons composed of stars in the field. To do so we have taken a composite spectrum from the Binary Population and Spectral Synthesis (BPASS) library v2.1 \citep{eldridge_binary_2017} and rescaled to a luminosity of 6 $\times$10$^{40}$ erg/s based on \citet{zhou_very_2022}. The stellar age has been matched to the age of the young association of stars \theulx\ seems to belong to \citep[20 Myr;][]{grise_ultraluminous_2008} while the metallicity is taken as for the gas. Figure~\ref{fig:heII_cloudy} (right panels) shows the results are broadly consistent, in line with our expectation that stars do not produce enough photons to excite \heii. The \texttt{phenomenological} model now predicts a level closer to the upper limit. Therefore we can conclusively say that the UV in \theulx\ is of the order of $\sim 1 \times 10^{39}$ erg/s and likely even lower than that. 

\section{Discussion}\label{sec:discussion}

While inclination effects have often been invoked to explain differences between hard and soft ULXs \citep{sutton_ultraluminous_2013}, such classification is clearly degenerate with mass-transfer rate \citep[e.g.][]{middleton_spectral-timing_2015, urquhart_optically_2016}. The mass-transfer rate affects the opening angle of the supercritical wind/funnel, and as a result many ULXs are observed to transit between spectral states \citep{feng_nature_2016, pinto_ultraluminous_2017, gurpide_discovery_2021}. The final appearance of a ULX is therefore determined by the relative position of our line of sight with respect to the opening angle of the funnel. Such degeneracy is difficult to break using line-of-sight observables. Instead, examining the ionising effects of the ULX in its environment, we can mitigate these uncertainties and access properties intrinsic to the source. 

Here we have attempted to do so by first using multi-band, state-resolved spectroscopy of the PULX \theulx\ to derive three realistic extrapolations of the SED into the UV. We have then examined the effects of each SED into their environment to attempt to put further constraints on the UV emission of this ULX. Based on realistic photo-ionization simulations of the surrounding bubble, we have shown (Section~\ref{sec:cloudy}) that the UV level in \theulx\ must be $\lesssim$ 1$\times$10$^{39}$ erg/s based on the non-detection of the \heii\ line. 

Note that on the basis of the inferred UV level we do not claim to have ruled out a particular accretion flow geometry e.g. the supercritical funnel described by the \texttt{sirf} (Section~\ref{sub:spectral_fitting}); we can only rule out the extrapolation presented in Section~\ref{sub:spectral_fitting}, but with the data at hand it may still be possible to find a description of the spectroscopic data compliant with the inferred UV level using the \texttt{sirf} model (or the \texttt{diskir}). To completely rule out the aforementioned models would require an in depth exploration of the parameter space, but such exploration of all the possible solutions is beyond the scope of this work. Nevertheless, our results suggest that regardless of the correct physical description of the accretion flow powering \theulx, it should predict a level of UV compliant with our estimate. Such correct extrapolation is obviously provided by our \texttt{phenomenological} model. Interestingly, this model has been interpreted as describing the accretion flow around a supercritically accreting magnetised NS \citep[e.g.][]{walton_evidence_2018, koliopanos_investigating_2019}, which may give credence to this model. Further constraints about the exact spectral shape would require more refined modelling of the photo-ionized region, which we leave for future work.

The inferred UV level in \theulx\ contrast with the extreme UV ($L_\mathrm{UV} \sim$10$^{40}$ erg/s) inferred in the ULX NGC~6946~X--1 through photo-ionized modelling from the surrounding nebula \citep{abolmasov_optical_2008} or directly observed in SS 433 \citep{dolan_ss_1997}\footnote{Although see \cite{waisberg_collimated_2019} for caveats related to the exact extinction towards SS 433 and spectral modelling.}, speculated to be an edge-on ULX \citep[e.g.][]{king_black_2002, fabrika_jets_2004}. As an additional test, we have run another set of simulations using the inferred blackbody in NGC~6946~X--1 by \citet{abolmasov_optical_2008} (see below Figure~\ref{fig:seds_ulxs}) and found that it would produce a \heii\ luminosity comparable to our best-fit \texttt{sirf} model (these are shown in Figure~\ref{fig:heII_cloudy} as gray lines). Therefore it is clear that the UV source in \theulx\ is \textit{intrinsically} dimmer than NGC~6946~X--1 (or SS 433). 

In passing, we note the bright UV source inferred by \citet{abolmasov_optical_2008} in NGC~6946~X--1 is in tension with the \texttt{BMC} model proposed by \citet{berghea_spitzer_2012}, which produces a UV emission of only $\approx$ 5$\times$10$^{38}$ erg/s (see below Figure~\ref{fig:seds_ulxs}). This model can only explain the broadband data invoking a rather bright ($L\sim 10^{39}$ erg/s) companion star and predicts a smaller shell than observed in the surrounding nebula, which may indeed point out to a brighter UV emission than inferred by \citet{berghea_spitzer_2012}. Together, this may favour the model reported by \citet{abolmasov_optical_2008}, but clearly a comprehensive study using \textit{both} optical and IR lines (with higher spatial resolution than afforded by \textit{Spitzer}) is needed to clarify this issue.

In the X-ray band, \theulx\ is also known to be harder than NGC~6946~X--1 \citep[e.g.][]{pintore_pulsator-like_2017, gurpide_long-term_2021}. The former is thought to be viewed through the optically thin supercritical disk/wind funnel, due to the low contribution from the soft component and presence of pulsations \citep{sathyaprakash_discovery_2019}. The latter instead is thought to be viewed through the wind \citep{middleton_broad_2014}. Here we have additional show that such differences translate into the UV too. If the differences in the X-ray band were solely due to a viewing angle effect, we would expect both systems to posses comparable levels of UV, provided the accretion flow geometry is also the same. Instead, our results suggest an intrinsic difference between these two sources, either due to the nature of the compact object or differences in the mass-transfer rate. Interestingly, the differences do not end there: the radio luminosity of the MF16 nebula surrounding NGC~6946~X--1 is among the brightest in the ULX family \citep{berghea_detection_2020}, whereas \theulx\ is radio quiet as pointed out by \citet{siwek_optical_2017}. A very similar source to \theulx\ in X-rays is Holmberg IX X--1 \citep{weng_evidence_2018}, suggested to host a NS based on its spectral X-ray hardness \citep{pintore_pulsator-like_2017}, even though pulsations in this source have not been found so far \citep{doroshenko_searching_2015, walton_evidence_2018}. A moderate UV luminosity of 10$^{39}$ erg/s has also been inferred in Holmberg IX X--1 from nebular observations \citep{abolmasov_kinematics_2008}. The source is also similarly surrounded by a radio quiet bubble, three orders of magnitude fainter than MF 16 \citep{berghea_detection_2020}. While the sample of ULXs with measured UV luminosities remains small, as alluded in \citet{gurpide_quasi_2024} it appears that the differences observed in the X-ray band cannot be solely due to an inclination effect. 

We note similar results were already noted by \citet{sutton_irradiated_2014}, who found higher reprocessing fractions were needed to explain the broadband spectra of soft ULXs when using a refined version of the disk irradiated model. This again indicates brighter optical/UV emission in soft ULXs, as implied here.

Considering the lower UV emission in \theulx\ and assuming the UV is mainly arising in the outflow photosphere, one possibility is to invoke a less powerful wind in \theulx. The softer spectrum of NGC~6946~X--1 could in principle be interpreted due to a higher mass-transfer rate \citep{narayan_spectra_2017} as opposed to an inclination difference, which would favour a higher ejection rate in this ULX compared to NGC~1313~X--2, hence accounting as well for the brighter UV luminosity in NGC~6946~X--1. 

Alternatively, a less powerful wind in \theulx\ could be ascribed, for instance, to the role of the magnetic field, which can reduce the radii over which the outflow is produced \citep[e.g.][]{vasilopoulos_ngc_2019}, thereby reducing outflow rates. Invoking an overall higher ejection rate in NGC~6946~X--1 would appear consistent with the brighter radio luminosity of MF16, but at present it is unclear whether the radio luminosities are solely due to outflow power differences, age or density of the surrounding medium. For instance, the expansion velocity in MF16 is $\approx$ 225 km/s \citep{dunne_what_2000}, which suggest a younger system compared to \theulx\ while the ISM density around NGC~6946~X--1 is also higher, by about a factor 10 \citep{dunne_what_2000, blair_hubble_2001,abolmasov_optical_2008}. A comparative study of NGC~6946~X--1 and \theulx\ using high-resolution spectroscopy in X-rays could help test some of these hypothesis.

\subsection{ULXs in the context of the nebular \heii\ line in metal-poor star-forming galaxies}

\begin{figure}
    \centering
    \includegraphics[width=0.49\textwidth]{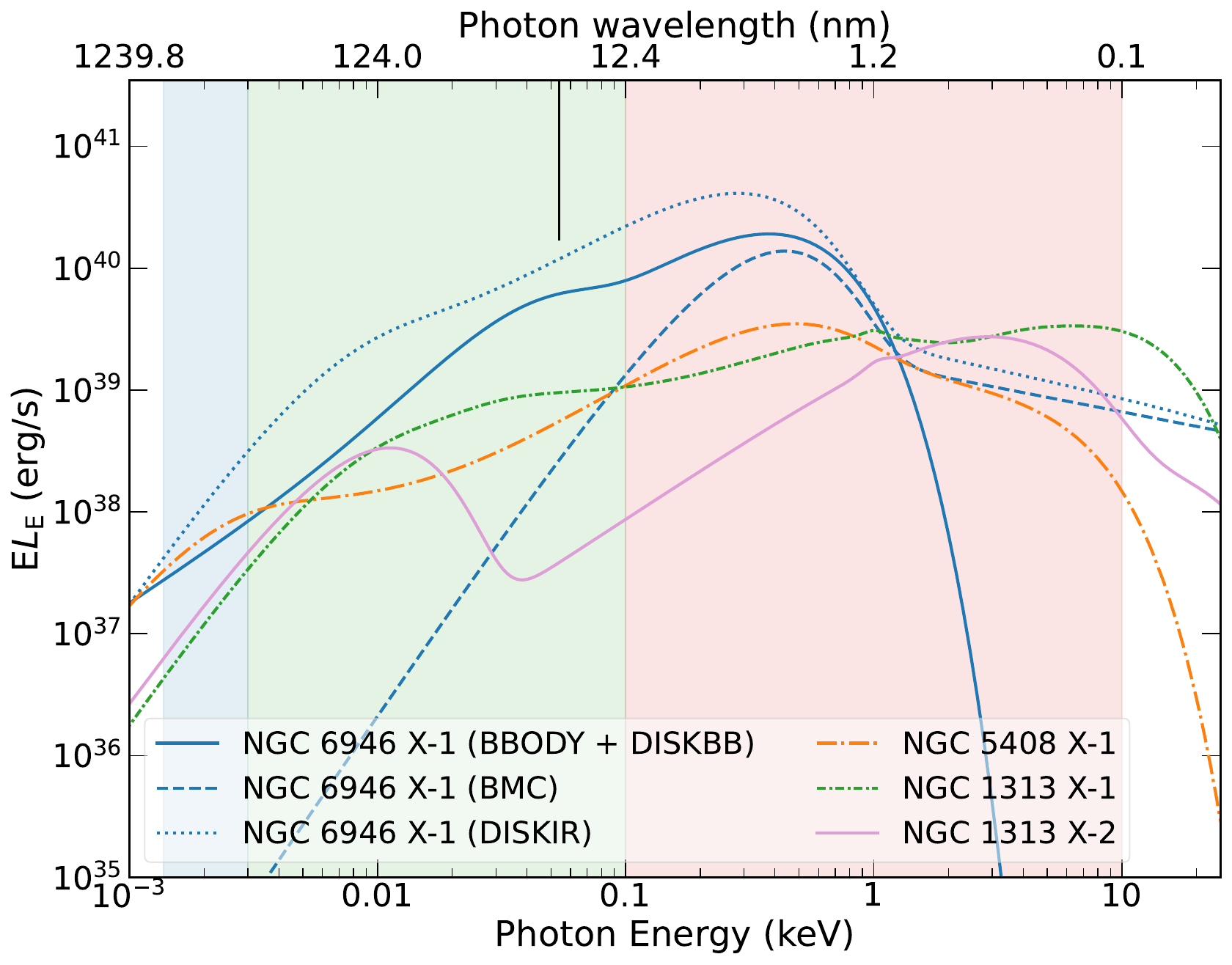}
    \caption{Comparison between the broadband SEDs derived for NGC~6946~X--1 \citep{abolmasov_optical_2008, berghea_spitzer_2012}, NGC~5408~X--1 \citep{kaaret_photoionized_2009}, NGC~1313~X--1 \citep{gurpide_quasi_2024} (\texttt{diskir} from the low state) and \theulx\ (this work). All SEDs are shown extinction corrected and re-scaled by their respective host galaxy distances. For the model from \citet{abolmasov_optical_2008} for NGC~6946~X--1 we have omitted the high-energy powerlaw. Based on photo-ionization modelling, \citet{berghea_spitzer_2012} found the \texttt{BMC} to be favoured over the \texttt{diskir} in NGC~6946~X--1 \citep[but see][]{abolmasov_optical_2008}. As per Figure~\ref{fig:broadband_seds}, a black vertical tick indicates the ionization potential of He$^{+}$.}
    \label{fig:seds_ulxs}
\end{figure}

As alluded in the Introduction, understanding whether ULXs could be behind the presence of \heii\ emission observed in the integrated spectrum of metal-poor star-forming galaxies remains uncertain \citep[e.g.][]{simmonds_can_2021, kovlakas_ionizing_2022}. Since most He+ ionising photons are produced in the EUV band, a major uncertainty affecting and highlighted in works attempting to ascertain whether ULXs could produce \heii\ in numbers compatibles with observations is the exact extrapolation of the ULX SED to the UV band. In Figure~\ref{fig:seds_ulxs} we compare our derived SED for \theulx\ (high state) with the broadband SEDs of the ULXs NGC~5408~X--1 \citep{kaaret_photoionized_2009}, NGC~6946~X--1 \citep{abolmasov_optical_2008, berghea_spitzer_2012} and NGC~1313~X--1 \citep{gurpide_quasi_2024}. These were derived using a combination of multi-band data and nebular observations similar to this work. All spectra are shown extinction corrected and re-scaled by the distances used in the quoted works. The SEDs of NGC~6946~X--1 (those derived by \citet{berghea_spitzer_2012}) and NGC~5408~X--1 were studied by \citet{simmonds_can_2021} in the context of the \heii\ problem. As a further comparison, \citet{simmonds_can_2021} estimated the number of He+ ionising photons ($Q$) per X-ray luminosity (0.5--8 keV in that work) from these spectra to be $\log (Q/L_\mathrm{X})$ = 9.61, 9.65 and 10.25 for NGC~5408~X--1 and the \texttt{BMC} and \texttt{DISKIR} models of NGC~6946~X--1, respectively. For the \texttt{blackbody} model inferred by \citet{abolmasov_optical_2008} \citep[using the \textit{Chandra} spectrum from][]{roberts_chandra_2003} we find $\log (Q/L_\mathrm{X}) \approx 10.1$ and for NGC~1313~X--1 we find 8.9--8.5 (for the high and low states, respectively). From our \texttt{phenomenological} model for \theulx\ we find $\log (Q/L_\mathrm{X}) = 8.5-8.8$ for the low and high states respectively. From the NGC~5408~X--1 and NGC~6946~X--1 models studied in \cite{simmonds_can_2021}, the authors found the \texttt{DISKIR} spectrum of {NGC~6946~X--1} in the Figure -- with the highest $Q/L_\mathrm{X}$ -- could produce \heii/H$\beta$ ratios compatible with observations, while the \texttt{BMC} required invoking rather high X-ray-to-galaxy contribution ratios \citep[see also][for a discussion on this aspect]{garofali_modeling_2023}. It is thus unlikely that sources such as \theulx\ could account for the observed line intensities in the integrated spectra of metal-poor star-forming galaxies. Moreover, given that \cite{berghea_spitzer_2012} found the \texttt{DISKIR} was not a good extrapolation to the UV based on the size and luminosity of the MF16 nebula surrounding NGC~6946~X--1, we see that rather extreme EUV luminosities need to be invoked to support ULXs as responsible for the presence of the \heii\ line in star-forming galaxies. 

In any instance, our work suggest there appears to be a wide range of ULX luminosities in terms of their capacity to excite He+. This may help explain why some of these star-forming galaxies do \textit{not} show strong \heii, as discussed also by \citet{simmonds_can_2021}. From this limited sample, it appears the UV luminosity correlates well with the soft X-ray luminosity (which is somewhat expected). If confirmed, this would imply that supersoft ULXs might even be more efficient in producing nebular \heii, but a statistical sample of ULXs with measured UV luminosities is needed to examine in detail the relationship between the X-ray and the UV emission in ULXs. 

\section{Conclusions} \label{sec:conclusion}

We have coupled multi-band spectroscopy of the PULX \theulx\ with upper limits on the \heii\ in its bubble nebula to constrain the level of UV/EUV emanating from this ULX. Based on photo-ionization modelling of the surrounding bubble nebula, we put an upper limit on its UV emission of the order of 1$\times$10$^{39}$ erg/s. The sample of ULXs with estimated UV emission remains small, but there appears to be a range of UV luminosities comparable to that observed in the X-rays (10$^{39-40}$ erg/s). Given these UV fluxes are \textit{intrisinc} to the source, we suggest such differences are incompatible with inclination effects and instead, may reveal differences in mass-transfer (and/or outflow) rate or in the nature of the accretor.

Based on the inferred UV in \theulx\, we consider it unlikely that ULXs like \theulx\ could produce enough He+ ionising photons to explain the presence of the \heii\ line in metal-poor, star-forming galaxies. However, as alluded in Section~\ref{sec:discussion}, there appears to be wide range of EUV luminosities existing in ULXs, which will dramatically influence their capacity to ionize He+. This may help explain why not all metal-poor star-forming galaxies show bright \heii.

We will present a more refined modelling of the nebula in a forthcoming publication, where we will put further constraints on the emission from the PULX \theulx. 

\section*{Acknowledgements}
We are grateful to the anonymous referee for their interest in our manuscript, and comments that helped improved it. We are also grateful to C. Berghea for help in reconstructing the broadband SEDs. N.C.S, A.G. and M. M. acknowledge support from the Science and Technology Facilities Council (STFC) consolidated grant ST/V001000/1. This research has made use of NASA/ESA Hubble Space Telescope obtained from the Hubble Legacy Archive, which is a collaboration between the Space Telescope Science Institute (STScI/NASA), the Space Telescope European Coordinating Facility (ST-ECF/ESA), and the Canadian Astronomy Data Centre (CADC/NRC/CSA), the NuSTAR Data Analysis Software (NuSTARDAS), jointly developed by the ASI Space Science Data Center (SSDC, Italy) and the California Institute of Technology (Caltech, USA) and the data supplied by the UK Swift Science Data Centre at the University of Leicester. Software used: Python v3.9.7, mpdaf \citep{piqueras_mpdaf_2017}, pyCloudy \citep{morisset_pycloudy_2013}, CAMEL \citep{epinat_massiv_2012}. 

\section*{Data Availability}

All the data used in this work can be found publicly in the corresponding archives. We have made available best-fit spectra, H$\alpha$ flux maps and some more ancillary data at \url{https://github.com/andresgur/data-gurpide-2024-NGC1313X2}.

\bibliographystyle{mnras}
\bibliography{references, refs}




\appendix

\section{\heii\ predictions for the low-state SEDs and impact of assuming collisional ionization equilibrium }
\begin{figure*}
    \centering
    \includegraphics[width=0.95\textwidth]{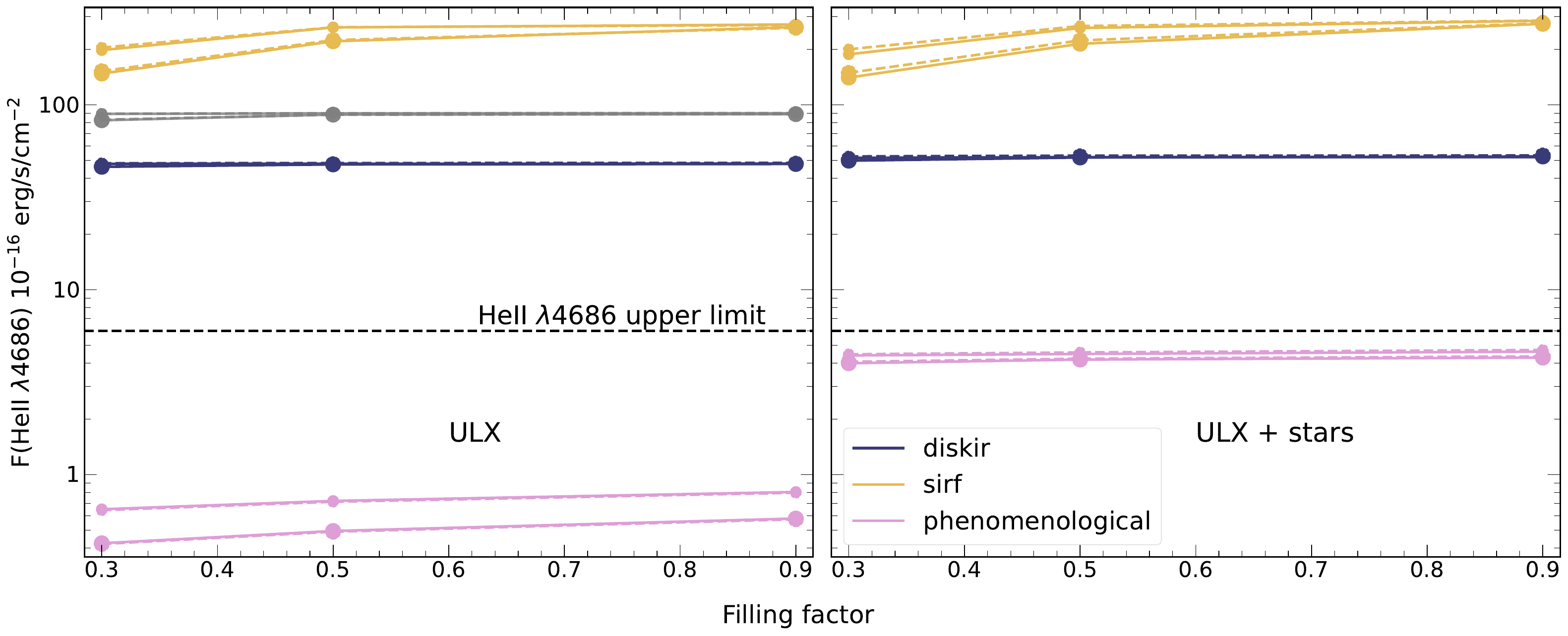}
    \caption{As per Figure~\ref{fig:heII_cloudy}, but now showing the predicted \heii\ fluxes for the low-state SEDs.} 
    \label{fig:heII_cloudy_appendix}
\end{figure*}

Figure~\ref{fig:heII_cloudy_appendix} shows \heii\ predictions as per Section~\ref{sec:cloudy} but for the low states (Figure~\ref{fig:broadband_seds_bbody}), which illustrates the fluxes are essentially consistent with those obtained for the high state, due to their similar UV luminosity. 

\begin{figure*}
    \centering
    \includegraphics[width=0.95\textwidth]{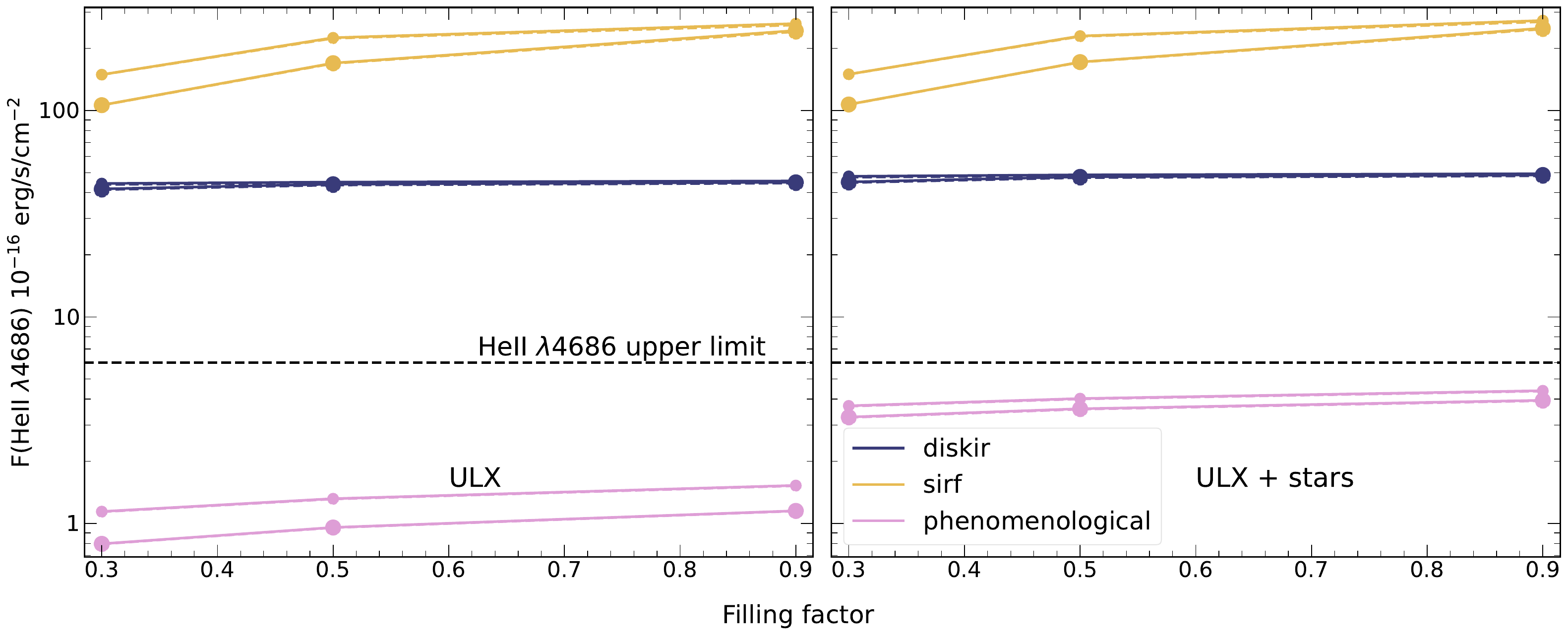}
    \caption{As per Figure~\ref{fig:heII_cloudy}, but now assuming the gas is in collisional ionization equilibrium with $T = $25,000 K.} 
    \label{fig:heII_cloudy_hot}
\end{figure*}

Figure~\ref{fig:heII_cloudy_hot} shows the effects of assuming the gas is in collisional ionization equilibrium (assuming $T = $25,000 K) (cf. Figure~\ref{fig:heII_cloudy}). The \heii\ predicted fluxes drop by less than $\sim$15\% for the \texttt{diskir} and \texttt{sirf} models. The most noticeable effect is found for the \texttt{phenomenological} when only the ULX is considered (left panel) where the predicted \heii\ flux doubles in some cases. These effects are in any instance small and do not affect our main conclusions.

\bsp	
\label{lastpage}
\end{document}